%% file: ms_pp.tex
\documentstyle[11pt,aaspp4]{article}
\newcommand{\etal}{et al. }
\newcommand{\eg}{e.g., }

\newcommand{\co}{\rm CO}
\newcommand{\oo}{{\rm O}_2}
\newcommand{\ac}{\rm C}
\newcommand{\io}{\rm O}
\newcommand{\hh}{{\rm H}_2}

\newcommand{\hi}{\rm HI}
\newcommand{\kps}{\,\textstyle\rm{km~s}^{-1}}
\newcommand{\lya}{Ly$\alpha$ }
\newcommand{\Mpc}{\,\textstyle\rm{Mpc}}
\newcommand{\coa}{CO(1$\rightarrow$0) }
\newcommand{\yr}{\,\textstyle\rm{yr}}
\newcommand{\Gyr}{\,\textstyle\rm{Gyr}}
\newcommand{\msun}{\,M_{\odot}}
\newcommand{\cm}{\,\textstyle\rm{cm}}
\newcommand{\GHz}{\,\textstyle\rm{GHz}}
\newcommand{\g}{\,\textstyle\rm{g}}
\newcommand{\mjy}{\,\textstyle\rm{mJy}}
\newcommand{\K}{\,\textstyle\rm{K}}

\lefthead{Frayer \& Brown}
\righthead{Evolution of CO, O$_2$, and Dust}

\begin{document}

\title{Evolution of the Abundance of CO, O$_2$, and Dust in the Early
  Universe}

\author{David T. Frayer} \affil{National Radio Astronomy
  Observatory\altaffilmark{1} and University of Toronto\\ Astronomy
  Department, University of Toronto, Toronto, Ontario, M5S
  3H8\\frayer@astro.utoronto.ca}

  \author{Robert L. Brown} \affil{National Radio Astronomy
    Observatory, Charlottesville, VA 22903\\rbrown@nrao.edu}

\altaffiltext{1}{The National Radio Astronomy Observatory is a facility
  of the National Science Foundation operated under cooperative
  agreement by Associated Universities, Inc.} 

\begin{abstract}

  We present a general set of calculations describing the chemical
  evolution of young massive galaxies and predict the evolution of the
  CO, O$_2$, and dust abundances as a function of age and metallicity.
  Over a wide range of input parameters, the models predict that (1) the
  total mass in gaseous metals peaks at early epochs ($z\sim1-3$) when
  approximately half the total baryonic mass is in stars and half is in
  gas, (2) at its extreme, the mass of gaseous metals ranges from a few
  to more than ten times larger than the mass of gaseous metals at the
  current epoch, (3) due to the larger O/C ratios in chemically young
  systems, the O$_2$/CO ratio may be of order unity within dark
  molecular clouds at early cosmological epochs, and (4) at early epochs
  the global volume--weighted O$_2$/CO abundance ratios corrected for
  photodissociation are lowered significantly compared to the dark cloud
  solutions due to the lower metallicities and the higher UV fields.
  For a variety of models, we calculate the evolution of the mass of CO,
  O$_2$, and dust.  We also compute the CO(1$\rightarrow$0) and
  O$_2(1,1\rightarrow 1,0)$ line intensities and the thermal dust
  emission as a function of redshift.  The model calculations suggest
  that both redshifted thermal dust emission and CO line emission will
  be easily observable in young massive galaxies with the next
  generation of millimeter and submillimeter wavelength telescopes.
  Although more challenging, the O$_2$ lines may be observable in
  chemically young galaxies.  Even though molecular oxygen has yet to be
  observed outside the solar system, and hence is much less abundant
  than CO in the Milky Way, the models indicate that O$_2$ may be
  significantly more abundant in the early stages of galaxy evolution.

\end{abstract}


\newpage
\section{Introduction}

Understanding the process of galaxy formation and the subsequent
evolution is one of the most active areas of astronomical research, but
our knowledge of the manner by which nascent gas forms stars and
galaxies is still seriously incomplete.  Since galaxies are expected to
evolve quickly during the first Gyr (\eg Tinsley 1972), observations of
young galaxies are crucial for our understanding of the star formation
and chemical evolutionary history of galaxies.  Optical observations
have revealed a plethora of information on young galaxies.  Recently, a
large population of star forming galaxies has been discovered at high
redshift (Steidel et al. 1996).  Studies of damped Ly$\alpha$ systems
(Wolfe et al. 1986; Lanzetta et al. 1991) suggest that the total amount
of neutral gas in these systems has decreased substantially from
$z\simeq 3.5$ to $z\sim0$, presumably due to the consumption of the gas
via star formation (Lanzetta et al.  1995).  Also, observations of the
damped Ly$\alpha$ systems indicate that significant amounts of metals
($Z\sim 0.1 Z_{\odot}$ at $z\sim2.5$, Pettini et al. 1994) and dust
(Pei, Fall \& Bechtold 1991; Fall \& Pei 1993) are produced at early
epochs.  In fact, observations of the gas near some high redshift QSOs
suggest abundances that are even in excess of solar (Petitjean, Rauch,
\& Carswell 1994; Turnshek et al. 1996).  All of these optical results
suggest that high redshift systems are already metal--enriched and at
least some are undergoing extensive star formation.

In addition to the optical results, recent high--redshift CO and dust
detections at mm and sub--mm wavelengths have confirmed the existence of
copious amounts of metals at very early epochs (Brown \& Vanden Bout
1991; Solomon, Downes, \& Radford 1992; Barvainis \etal 1994a; McMahon
\etal 1994; Chini \& Kr\"{u}gel 1994; Dunlap \etal 1994; Isaak \etal
1994; Ivison 1995; Ohta et al. 1996; Omont et al. 1996a; Omont et al.
1996b).  These CO and dust observations have the advantage of providing
direct measurements of the mass of the gaseous reservoir from which the
stars form.  Such observations are crucial for constraining the
evolutionary models for galaxies.  Only by combining both optical and
radio observations will we gain a complete picture of galaxy formation
and evolution.

Despite the cumulative expenditure of many thousands of hours of
telescope time by several groups in quest of high redshift CO and dust
detections, there has yet to be a complementary investigation into the
expected quantity and composition of the molecular gas and dust in young
galaxies.  In this paper we address these issues by studying the
chemical evolution of the gas in massive galaxies undergoing large rates
of star formation at early epochs, and we predict the evolution of the
molecular gas and dust content.

In addition to the aforementioned CO and dust detections at high
redshift, this theoretical investigation is motivated by several other
observational results.  From observations of nearby galaxies (Wilson
1995; Arimoto, Sofue, \& Tsujimoto 1996), we are beginning to understand
how metallicity affects the molecular abundances in chemically young
galaxies.  Furthermore, observations of nearby metal--poor galaxies with
the Hubble Space Telescope (HST) have shown that the C/O ratio increases
with metallicity (Garnett \etal 1995).  This result has important
ramifications on the molecular composition of chemically young systems.
In particular, the O$_2$/CO ratio decreases exponentially with
increasing C/O ratios (Langer \& Graedel 1989 [LG89]).  One of the major
goals of this paper is to investigate how variations of the C/O ratio
affect the O$_2$/CO ratio at early cosmological epochs.

In order to calculate the chemical evolution of the gas in young
galaxies, we formulate a numerical model (\S2) based on the standard
equations governing the chemical evolution of stars and gas in galaxies
(Tinsley 1980).  The solutions to these equations depend sensitively on
(1) the initial mass function (IMF), which describes the fraction of
stars formed at each initial stellar mass, (2) the stellar yields, (3)
the parameters governing the star formation rate (SFR), and (4) the time
scales for which gaseous systems collapse to form galaxies.  The redshift
corresponding to each time scale depends, of course, on the cosmological
parameters.  For all redshift scales in this paper, we adopt the
standard Friedmann cosmology with $q_o=1/2$ and $\Lambda=0$.  In such a
universe, the age of a galaxy as a function of redshift is
\begin{equation}
T(z)=\frac{2}{3 H_o}\left[ \left(\frac{1}{1+z}\right)^{3/2} -
\left(\frac{1}{1+z_f}\right)^{3/2} \right],
\end{equation}
where $z$ is the redshift of the galaxy and $z_f$ is the redshift at
which the galaxy forms (\eg Fritze-v. Alvensleben \etal 1989).  In this
paper we use $z_f=5$ (the redshift of the most distant known QSO;
Schneider, Schmidt, \& Gunn 1991) and $H_o=80\kps \Mpc^{-1}$ (Pierce
\etal 1994; Freedman \etal 1994).

The stellar yields, which quantify the amount of gaseous material
returned into the interstellar medium (ISM) via supernovae, stellar
winds, and planetary nebulae, are very important input parameters into
the chemical evolutionary models.  We use the stellar yields calculated
as a function of metallicity for massive stars by Maeder (1992) [M92]
and Woosley \& Weaver (1995) [WW95].  The stellar yields are very
dependent on supernova energies, mass loss via winds (M92), the masses
of stellar remnants, and the parameters governing stellar
nucleosynthesis, such as the mixing length parameter (Renzini \& Voli
1981 [RV81]).
  
Since the properties of protogalaxies are still very uncertain, we
develop a general set of models.  By varying the above mentioned
parameters, we calculate the gas mass, stellar mass, and metallicity as
a function of age for 24 different models (\S3).  The models are loosely
constrained by the observed properties of the Milky Way and by the
abundances of nearby {\sc Hii} galaxies.  We follow the evolution of the
oxygen and carbon elemental species as well as the total metallicity (Z)
since the molecular and dust abundances are sensitive to these
quantities.

By combining the results of the chemical evolutionary models with
previous detailed computations of gas--phase molecular chemistry (LG89;
Millar \& Herbst 1990 [MH90]), we predict the evolution of the O$_2$/CO
ratio (\S4).  We discuss several scenarios explaining the apparent lack
of O$_2$ in the local universe, and consider the implications for
galaxies at earlier epochs.  We calculate the evolution of the total
mass contained within the CO, O$_2$, and dust species.  These
calculations are sensitive to several uncertain parameters, such as (1)
the fraction of gas in molecular form, (2) the fraction of molecular gas
in dark cloud cores ($A_{V}\ga5)$, (3) the evolution of the
$N(\co)/N(\hh)$ ratio, (4) the $M(\hh)$ to $L(\co)$ conversion factor as
a function of metallicity and environment, (5) the effect that
photodissociation has on the extent of the CO and O$_2$ emission
regions, (6) the effect that turbulent mixing has on the molecular
chemistry, (7) the relative depletion and subsequent chemistry of atomic
and molecular species on grains, and (8) variations of the gas--to--dust
ratio with metallicity.  In addition, the models are calibrated using
assumptions based on the properties and chemistry within Galactic clouds
which may not necessarily be applicable for intense starburst systems.
Given the uncertainties in these parameters, we present several
scenarios comparing the dust and molecular gas content of the current
epoch with that of the early universe.  In \S5 we compute the evolution
of the intensities of the redshifted \coa and O$_2(1,1\rightarrow 1,0)$
lines and the thermal dust continuum emission.  Finally, we discuss the
feasibility of observing these species at high redshift using current
and future radio telescopes.

\section{Chemical Evolution Model}
\label{sec-chemev}
\subsection{Basic Equations}
\label{sec-basiceqn}
We use the formalism of Tinsley (1980) to solve the standard equations
governing the evolution of gas and stars in galaxies.  At all times the
total mass in the system ($M_T$) is the sum of the gas mass $M_g$, the
stellar mass $M_s$, and the total mass contained in stellar remnants
$M_r$.  These quantities are individually determined by solving the
following set of time--dependent differential equations:
\begin{equation}
\label{eq:mtt}
\frac{dM_T}{dt} = f(t),
\end{equation}
\begin{equation}
\label{eq:msrt}
\frac{d(M_s+M_r)}{dt}= \Psi(t) - E(t),
\end{equation}
\begin{equation}
\label{eq:mgt}
\frac{dM_g}{dt}= -\Psi(t) + E(t) + f(t).
\end{equation}
The function $f(t)$ describes the net flow of material into or out of the
system.  In the case of inflow $f(t)$ is positive, and with net outflow
$f(t)<0$.  The total consumption rate of the gas by star formation is
represented by $-\Psi(t)$.  The rate at which material is ejected back
into the system from stars via supernovae, planetary nebulae, and
stellar winds is
\begin{equation}
\label{eq:et}
E(t)= \int_{m_t}^{m_u} (m-w_m)\Psi(t-\tau_m)\Phi(m)dm,
\end{equation}
where $w_m$ is the remnant mass of a star with an initial mass $m$,
$\Phi$ is the IMF ($\Phi = dN/dm)$, and $\tau_m$ is the lifetime of a
star of mass $m$.  The product $\Psi(t-\tau_m)\Phi(m)$ is the total
number of stars of mass $m$ formed at the time of their birth
$(t-\tau_m)$.  The lower limit of integration $m_t$ is the stellar mass
for which $\tau_m = t$, and $m_u$ is the upper limit on the mass of a
star.  Equation~(\ref{eq:et}) uses the standard approximation that each
star undergoes its entire mass loss at the end of its lifetime. Since
this approximation fails on short time scales due to mass loss from
stellar winds, we restrict analyses to time scales $t>10^{6} \yr$.  The
production rate of stellar remnants is described by
\begin{equation}
\label{eq:mrt}
\frac{dM_r}{dt}= \int_{m_t}^{m_u} w_m\Psi(t-\tau_m)\Phi(m)dm.
\end{equation}
The rate of change of stellar mass $dM_s/dt$ is solved directly by combining
equation~(\ref{eq:msrt}) and equation~(\ref{eq:mrt}).

Analogous to equation~(\ref{eq:mgt}), the time--dependent metal
abundance $Z_i$ of species $i$ is represented by
\begin{equation}
\label{eq:zmgt}
\frac{d(Z_iM_g)}{dt}=-Z_i\Psi(t) + E_{Z_i}(t) + Z_{i_f}f(t),
\end{equation}
where $Z_{i_f}$ is the metallicity of the infalling or outflowing gas.  The
ejection rate of metals is 
\begin{equation}
E_{Z_i}(t)= \int_{m_t}^{m_u} [(m-w_m)Z_i(t-\tau_m) +
mp_{z_i}(m)]\Psi(t-\tau_m)\Phi(m)dm, 
\end{equation}
where $mp_{z_i}(m)$ is the stellar yield of metal $Z_i$ from a star with
initial mass $m$.  In the above equation, we have made the correction
suggested by M92 which removes a term from the formula given by Tinsley
(1980) in order to conserve overall mass.  This minor correction does
not modify the widely used analytical solutions given by Tinsley.

\subsection{Analytical Solutions}
\label{sec-analy}
The instantaneous recycling approximation is often used to remove the
time dependency from the differential equations in \S\ref{sec-basiceqn}
which allow for simple analytical solutions.  In this approximation
stars above a certain mass value (\eg $\sim 10 \msun)$ die immediately
after birth and stars with masses less than this value live forever, so
that $(t-\tau_m)$ can be replaced simply by $t$.  In the case of the
Closed Box Model, where $f(t)=0$, and assuming that $M_g=M_T$ and $Z=0$
at $t=0$, it is straight forward to show (\eg Tinsley 1980)
\begin{equation}
\label{eq:cbox}
Z_i=y_i\ln(\mu^{-1}),
\end{equation}
where $\mu$ is the gas fraction $M_g/M_T$.  The net yield $y_i$ is the
mass of new metals of species $i$ ejected by all stars per unit mass
that is locked into stars and stellar remnants; it is given by
\begin{equation}
\label{eq:nety}
y_i=\frac{1}{(1-R)}\frac{\int_{m_l}^{m_u}mp_{z_i}\Phi(m)
dm}{\int_{m_l}^{m_u} m\Phi(m)dm},
\end{equation}
where the IMF is defined between the upper ($m_u$) and lower ($m_l$)
stellar mass limits.  The returned fraction $R$ is the fraction of
original stellar mass that has been ejected back into the system and is
expressed by
\begin{equation}
R=\frac{\int_{m_l}^{m_u}[m-w_{m}]\Phi(m)
dm}{\int_{m_l}^{m_u} m\Phi(m)dm}.
\end{equation}
Traditionally, $R$ and $y_{z}$ have been the main parameters for studying
galactic chemical evolution models since they depend on the adopted IMF
and the stellar yields predicted by stellar evolution, but not on $\Psi(t)$.

Another popular analytical solution is the case of simple infall.  The
simple infall model assumes that the star formation rate is balanced
with the rate of infall plus stellar gas loss so that the total {\it
  gas} mass remains constant with time.  When instantaneous recycling is
also assumed, the equations have the solution (Larson 1972)
\begin{equation}
\label{eq:sinfall}
Z_i=y_i[1-e^{(1-\mu^{-1})}].
\end{equation}
Other more complicated infall solutions can be found in
Clayton (1987).  Solutions including the effects of galactic winds
($f(t) > 0$) are discussed by Matteucci \& Chiosi (1983), and a general
set of equations representing both gaseous inflow and outflow is given
by Edmunds (1990).  

The instantaneous recycling approximation, which permits these simple
analytical solutions, is generally not valid in cases where the SFR is
strongly time dependent since $\Psi(t-\tau_m) \neq \Psi(t)$.  Also, the
time--dependent ratios of metal abundances are determined incorrectly
when ignoring stellar lifetimes since heavy elements are produced in
different quantities in stars with different lifetimes (i.e., different
masses).  Equally important, it is impossible to derive information on
the earliest stages of evolution since the approximation is only valid
when studying effects which vary on time scales much larger than the
lifetimes of massive stars ($\gg 10^{7} \yr)$.

\subsection{Numerical Solution}

We numerically solve the equations of \S\ref{sec-basiceqn} to avoid the
necessity of using the instantaneous recycling approximation.  We follow
the evolution of the gas using a time step of $2\times 10^{6} \yr$.
Numerical solutions of the basic evolutionary equations are very general
and have been used extensively to investigate many astrophysical
questions (\eg Matteucci \& Tornamb\`{e} 1987; Rana 1991; Timmes,
Woosley, \& Weaver 1995).  In this paper we solve these equations in
order to study the evolution of carbon, oxygen, and total metallicity in
young galaxies.  Since the characteristics of young galaxies are
uncertain, we vary the input parameters to understand how the results
are modified with changing inputs.  In Table~1 we summarize the values
of the input parameters for 24 different models, covering the range of
likely astrophysical parameters.  We discuss particular details of these
parameters in the following subsections.

\subsubsection{The IMF, the Star Formation Rate, and Infall Parameters}

The solutions to the equations governing the chemical evolution of galactic
systems depend sensitively on the adopted IMF, the star
formation history, and the rate of gaseous inflow or outflow.  We adopt
a single slope power--law approximation to the IMF of the form 
\begin{equation}
\Phi(m)\propto m^{-(1+x)},
\end{equation}
where $x$ is the slope of the IMF.  We compare the results of a Salpeter
power law ($x=1.35$, Salpeter 1955) with those of a steeper ($x=1.7$)
IMF and a flatter IMF ($x=1.1$).  In the equations of
\S\ref{sec-basiceqn}, the IMF is normalized such that
\begin{equation}
\int_{0.08\msun}^{m_{u}} m\Phi(m)dm = 1,
\end{equation}
where $m_u$ is the upper limit to the mass of a star.  The calculations
of WW95 include stars as massive as $m_u = 40\msun$, while the
calculations of M92 include stars with mass up to $m_u = 120\msun$.
 
To test the dependency on the star formation history, we use a
simple power law for the star formation rate of the form (Schmidt 1959)
\begin{equation}
\label{eq:sfr}
\Psi(t)= \nu M_T(t)[\mu(t)]^{n}.
\end{equation}
The exponent is varied such that the dependence on the gas fraction is
between linear and quadratic; we use the values of $n=1.0$, 1.5, and
2.0.  In studying of the evolution of elliptical galaxies, Matteucci \&
Tornamb\'{e} (1987) adopt $n=1$.  Alternatively, for spiral galaxies
Dopita \& Ryder (1994) find that the values of $n\simeq 1.5-2$ are
consistent with observations.  We vary the efficiency factor between the
values $\nu =$~0.5, 2, and 5~Gyr$^{-1}$.  In studying Galactic chemical
evolution, Timmes \etal (1995) find that $\nu=2.8$~Gyr$^{-1}$ is the
best fit to the observational data.  Even though surface densities may
be more directly related to the SFR (e.g., in terms of the Toomre
stability criterion for disks; Kennicutt 1989), we use total masses
instead of mass surface densities in equation ~(\ref{eq:sfr}).  The
advantage of using this formalism is that we do not need to adopt a
particular dynamical model or geometry (e.g., disk or spheroid).  By
using total masses, we only determine the total global SFR.  Granted,
this approximation over simplifies the actual picture of star formation
in galaxies which may differ considerably from ellipticals and spirals
and undoubtedly includes interactions and mergers, but the approximation
is suitable for studying global properties over long time scales
($t\ga$~Gyr).  For example, in the Galactic disk where $M_T \sim
10^{11}\msun$ and $\mu=0.1$ (Rana \& Basu 1992), we find that
equation~(\ref{eq:sfr}) implies that the $SFR\sim 6\msun \yr^{-1}$,
using $n=1.5$ and $\nu=2$~Gyr$^{-1}$.  This value compares well with the
Galactic estimate of the $SFR\simeq5\msun\yr^{-1}$ (\eg Mezger 1987).
Extrapolating back to earlier epochs where $\mu$ could be of order
unity, the most massive galaxies ($M_T\sim 10^{12}\msun$) are expected
to have total star formation rates of order $10^{3}\msun \yr^{-1}$.

We study the numerical solutions for both the closed box models
($f(t)=0$) and infall models ($f(t)>0$).  For the infall models, we
assume that the initial mass is zero and that the accreting gas is
primordial ($Z_{i_f}=0$ in eq.~[\ref{eq:zmgt}]).  To describe the infall
rate, we use an exponential function
\begin{equation}
f(t)=\frac{M_T(t\rightarrow \infty)}{\tau_f}\exp(-t/\tau_f),
\end{equation}
where $M_T(t\rightarrow \infty)$ is the maximum mass of a system as
$t\rightarrow \infty$ and the time scale for formation is $\tau_f$.  We
have chosen $M_T(t\rightarrow \infty) = 10^{12} \msun$ since massive
galaxies are more readily observable at high redshift.  For galaxies
with smaller masses, the results presented in this paper may be scaled
accordingly.  In the formation of a thin disk, Burkert, Truran, \&
Hensler (1992) find that $\tau_f\simeq 4$~Gyr.  Spheroidal systems are
expected to form on shorter time scales ($\tau_f\sim 1$~Gyr; Theis,
Burkert, \& Hensler 1992).  In this paper, we vary $\tau_f$ between the
values of 1, 2, and 4~Gyr.  For the limiting cases of $\tau_f\rightarrow
0$~or~$\tau_f\rightarrow \infty$, we recover the closed box model
solution ($f(t)=0$).

\subsubsection{Stellar Yields and Lifetimes}
\label{sec-yields}
Stellar yields are arguably the most important input parameters
governing the evolution of elemental abundances.  In order to include
the dependency of the stellar yields as a function of metallicity, we
adopt the yields from massive stars based on the calculations of WW95
and M92.  Woosley and Weaver determined the ejected masses resulting
from Type II supernova models of varying initial stellar masses ($12
\leq M/\msun \leq 40$) and varying initial metallicities (0, $10^{-4}$,
0.01, 0.1, and 1.0 $Z_{\odot}$).  Their calculations do not include
pre--supernova mass loss.  The ejected yields in massive stars depend on
the energy of the supernova explosion.  Simply stated, energetic
explosions return more material to the ISM than less energetic
explosions which tend to leave more massive stellar remnants.  We adopt
the yields from the models using intermediate (Case B) and large (Case
C) supernova energies (WW95).  For the purposes of our calculations, we
linearly interpolate the stellar yields and the masses of the stellar
remnants for intermediate metallicities.  For metallicities larger than
solar, we use the yields calculated for solar metallicity.

Maeder's models, on the other hand, do include mass loss by stellar
winds prior to the supernova event.  To assess these effects, we use the
yields and remnant masses of M92.  Maeder has shown that mass loss in
massive stars leads to stellar yields that are highly sensitive to the
initial metallicity.  For massive stars ($M>25\msun$) the oxygen yields
are expected to decrease with metallicity due to the winds, whereas the
carbon yields increase with metallicity.  These tendencies are expected
since massive stars at high metallicity have sufficiently high
opacities, such that large amounts of helium and carbon are ejected into
the ISM before being processed into heavier elements.  This produces a
low oxygen yield and a high carbon yield.  The reverse is expected at
low metallicities.  Since at low metallicities the opacity is low and
mass loss is less important, most of the new helium and carbon can be
processed further into heavier elements prior to the supernova event.
The resulting ejecta have a high oxygen yield and a low carbon yield.
Unlike the yields of WW95 which cover a complete grid of metallicities,
the yields of M92 are only for $Z=0.001$ and $Z=0.02\simeq Z_{\odot}$.
We avoid extrapolations and use the yields at $Z=0.001$ for $Z\leq
0.001$, the yields at $Z=0.02$ for $Z\geq 0.02$, and linearly
interpolate for intermediate metallicities.  As pointed out by Prantzos,
Vangioni--Flam, \& Chauveau (1994), the stellar yields for $Z<0.001$ are
expected to be similar to those at $Z=0.001$ since mass loss is already
negligible at this metallicity.  

In Figure~\ref{yields} we show the expected stellar yields for massive
stars from the work of WW95C (Case C) and M92 at solar ($Z_{\odot} =
0.02$) and subsolar metallicities.  At low metallicities, the yields
of both M92 and WW95C are similar.  The ratio of the oxygen to carbon
yields increases as a function of stellar mass, a simple reflection of
the fact that more massive stars burn more carbon into oxygen than less
massive stars.  Since massive stars have the shortest lifetimes, these
results indicate that we should expect an excess of oxygen relative to
carbon at the earliest epochs.  At high metallicities,
Figure~\ref{yields} shows how the inclusion of winds substantially
modifies the resulting yields.  Without stellar winds massive stars have
yields of $mp_O/mp_C > 1$, whereas if winds are included then
metal--rich massive stars ($M >25 \msun$) become an important source of
carbon ($mp_O/mp_C < 1$; M92).  Therefore, the stellar winds in massive
stars tend to amplify the carbon abundance relative to oxygen as a
function of increasing age and metallicity.

For low and intermediate mass stars ($1\leq M/\msun \leq 8$), we use the
stellar yields from RV81.  We use the yields predicted with a mass loss
parameter of $\eta =0.33$ and with the mixing length (``hot--bottom''
burning) parameters of $\alpha =1.5$ and $\alpha =0$ given for $Z=0.02$
and $Z=0.004$ (RV81).  As with the yields of M92, we linearly
interpolate for intermediate metallicities.  We use the yields and
remnant masses at $Z=0.004$ for $Z\leq 0.004$ and the yields and remnant
masses at $Z=0.02$ for $Z\geq 0.02$.  The relative yields of C, O, and N
are extremely sensitive to the mixing length parameter for intermediate
mass stars ($4 \la M/\msun \la 8$).  In the case of $\alpha=0$ (no
``hot--bottom'' burning) the carbon yield dominates, whereas for
$\alpha=1.5$ nitrogen production is important.  Correspondingly, the
mixing length parameter affects the epoch at which intermediate mass
stars are significant sources of carbon enrichment.  For $\alpha=1.5$,
only stars less massive than $5\msun$ are expected to contribute
significantly to the carbon yields, suggesting a relevant time scale of
$\sim 10^{8} \yr$.  For $\alpha =0$, however, the more massive
intermediate mass stars are an important carbon source, decreasing the
relevant time scale to only $\sim 10^{7} \yr$.  Recent work by Chieffi,
Straniero, \& Salaris (1995) suggest that the mixing length parameter is
$\alpha\simeq1.6$ for low metallicity stars ($Z\sim 0.01 Z_{\odot}$) and
increases to $\alpha\simeq1.9$ for stars of intermediate metallicity.
In this paper we concentrate on the models using $\alpha=1.5$ and do not
include the possible increase of $\alpha$ with metallicity.

We have ignored the contribution of Type I supernovae to the stellar yields
since Type~I events produce mostly Fe and Ni species and only a little
carbon and oxygen.  More importantly, Timmes \etal (1995) have
shown that the Type I contribution is relatively small compared to
yields from single star evolution.  Using the amplitude of Type~I
contributions as a free parameter and comparing with Galactic
abundance observations, they estimate that the Type~I contribution is
only 0.007 times that of the contribution from single stars over the
mass range of $3\msun$ to $16\msun$ (the minimum and maximum binary mass
which produces a carbon--oxygen white dwarf).

We use the stellar lifetimes from Theis \etal (1992):
\begin{equation}
\tau(m)=\left\{  \begin{array}{ll} 1.2\times10^{10}
(\frac{m}{\msun})^{-2.78} \yr & m < 10 \msun \\
1.1\times10^{8} (\frac{m}{\msun})^{-0.75} \yr & m \geq 10 \msun. \end{array}
\right.
\end{equation}
As with the stellar yields, the stellar lifetimes also vary with
metallicity.  However, these differences are most important for low mass
stars.  For low mass stars at high metallicity, the opacities are larger
so the luminosities are lower, hence, producing longer lifetimes.  For a
$3\msun$ star the lifetime differences are about 20\% between the cases
of $Z=0.001$ and $Z=0.02$, and for $M< 1\msun$ the differences are
expected to be larger than 50\% (Schaller \etal 1992).  Since we are
primarily concerned with massive stars at early times, our main results
are insensitive to these differences.

\section{Results of Chemical Evolution}
\label{sec-chem_results}
The parameters of Table~1 were varied around the intermediate values of
$x=1.35$, $\nu=2$, and $n=1.5$ which represent typical values for
galaxies at the current epoch.  We have studied models using the yields
of WW95 and M92 to test the importance of stellar winds.  In
Figure~\ref{evolt1} we show the evolution of gas mass, stellar mass,
total mass, and metallicity as a function of age for a typical closed
box model, while Figure~\ref{evolt11} displays the same quantities for
an infall model.  Since we are studying global properties, the derived
metallicities represent a mass weighted average for the entire galactic
system.  For the Galactic disk, which has a metallicity gradient that
ranges from $Z \sim Z_{\odot}$ in the solar neighborhood to $Z\sim 5
Z_{\odot}$ in the central regions (Pagel \& Edmunds 1981), the average
metallicity is larger than solar.

In Figure~\ref{zmgm92} we present the evolution of gaseous metals $Z\mu$
as a function of age for several different parameters governing the SFR.
For the parameters chosen in these models, the peak of gaseous metals as
a fraction of the total mass occurs fairly rapidly, generally within
$\sim$ 2~Gyr.  After which, the amount of gaseous metals decreases as
the gas is consumed by the ongoing star formation and locked into low
mass stars and stellar remnants.  The consumption rate of the gas is, of
course, very sensitive to the adopted star formation rates.  For $n>1$
the SFR decreases faster than the gas fraction, permitting a slow
decline in the fractional amount of gaseous metals.  Similarly,
decreasing $\nu$ also decreases the SFR and reduces the consumption rate
of gaseous metals.  This parameterization of the SFR could be used to
represent galaxies of different morphological types.  Early--type
galaxies, such as ellipticals and S0 galaxies, currently have little gas,
suggesting high $\nu$ and low $n$.  Alternatively, gas--rich, late--type
spiral galaxies could have low $\nu$ and high $n$ (Sandage 1986).

In Figure~\ref{muzm92} we plot the mass of gaseous metals as a function
of the gas fraction $\mu$.  The SFR parameters have little effect on the
shape or maximum value of the curves, but they do determine the time
scale for which a system will evolve to a given $M_g/M_T$.  Although the
shape of the curves is also basically independent of the IMF, the
maximum amount of gaseous metals is extremely sensitive to the slope of
the IMF.  For a steep IMF, there are fewer massive stars so less metals
are produced.  Alternatively for a shallow IMF, more metals are
produced.  Since the maximum amount of gaseous metals is relatively
insensitive to the SFR parameters, the IMF for these models can be
constrained by comparing the implied net yields with previous studies.
For the analytical solution of the closed box model
(eq.~[\ref{eq:cbox}]), the maximum amount of gaseous metals occurs at
$\mu=1/e=0.368$ (Bally, Shull, \& Hamilton 1993) and the implied net
yield is $2.72(Z\mu)_{max}$.  For an IMF with $x=1.35$ (Run~1), the
implied net yield is $y_z=0.025$, which is well within the range of
values estimated for our Galaxy: $y_z =0.036$ (Pagel 1987) and
$y_z=0.011$ (Rana 1991).  For $x=1.1$ (Run~2) and $x=1.7$ (Run~3), the
implied net yields are outside the range of acceptable Galactic
estimates (Table~2), suggesting that the Salpeter IMF is an adequate
approximation for describing the evolution of the Galaxy.

In Figure~\ref{paul} we compare the numerical solutions to the
analytical solutions given in \S\ref{sec-analy}.  The analytical
solution of the closed box model (eq.~[\ref{eq:cbox}]) was plotted using
a net yield of $y_z=0.025$, which is the implied net yield for the
closed box numerical solution of Run~1.  For the infall models, the
maximum $Z\mu$ occurs at $\mu\simeq 0.47$, which is at an earlier age
than the maximum $Z\mu$ of the closed box models.  For the analytical
solution of the simple infall model (eq.~[\ref{eq:sinfall}]), we find
that the implied net yield is $3.15(Z\mu)_{max}$, which corresponds to
$y_z=0.024$ for the numerical infall solution of Run~11.  The
consistency between the analytical and numerical solutions in
Figure~\ref{paul} is somewhat remarkable considering that the actual net
yield for the numerical models changes with metallicity as the
individual stellar yields change (eq.~[\ref{eq:nety}]), whereas for the
analytical solutions the net yield is assumed to be constant.  As
expected, when the infall rate $f(t)\rightarrow 0$ in the numerical
infall model, the solution deviates from that of the analytical infall
model and approaches the solutions of the closed box models.  The
primary significance of these results on the evolution of gaseous metals
is straightforward.  Basically independent of the details of the model,
the total mass of metals in the ISM of galactic systems is predicted to
peak at early epochs ($z\sim 1-3$) when the total gas mass is similar to
the total stellar mass.

Although understanding the evolution of metallicity is important,
comprehending the evolution of the C/O ratio is more interesting and
important for this study, since the molecular abundances are very
sensitive to this ratio.  As discussed in \S\ref{sec-yields},
low--metallicity massive stars produce more oxygen than carbon.  Since
the bulk of the carbon production occurs at later epochs, either from
intermediate and low mass stars and/or from the winds of metal--rich
massive stars, the gaseous C/O ratio is expected to increase with
metallicity.  Abundance determinations of Galactic stars are consistent
with these expectations.  In addition, extragalactic observations with
the HST have provided similar results for H{\sc ii} galaxies (Garnett
\etal 1995).  In Figure~\ref{coz} we plot the observed C/O abundance
ratios along with the solutions of several models calculated for a
variety of input parameters.  The metallicities of the stars observed by
Tomkin \etal (1992) and Clegg, Lambert, \& Tomkin (1981) are determined
from their Fe abundances, while the metallicities of the extragalactic
sources observed by Garnett \etal (1995) are determined from the oxygen
abundances, assuming the solar relationship of $Z=23(\io/{\rm H})$
(Grevesse \& Anders 1989).  At the low metallicities, the solutions
based on the yields with the largest supernova energies (M92 and WW95C)
better match the observations than the case of lower supernova energies
(WW95B).  The M92 solution with $m_u=40\msun$ is similar to the WW95C
solution which also uses $m_u=40\msun$.  With the M92 models, changing
the mixing length parameter from $\alpha=0$ to $\alpha=1.5$ has little
effect, indicating that most of the carbon is produced by the stellar
winds from massive stars with $Z\ga 0.1 Z_{\odot}$.  Previously,
Prantzos \etal (1994) have accounted for the rise of the C/O ratio in
the disk of our Galaxy using the yields of M92 without requiring a
significant contribution from intermediate and low mass stars.  Using
calculations based on the yields of WW95, however, the resulting
abundances are sensitive to $\alpha$ because in these models the
intermediate mass stars are the primary source of carbon.

In Figure~\ref{cozm} we further demonstrate the dependency of
the C/O ratio as a function of metallicity on other parameters such as
the adopted IMF, star formation history, and infall rate.  Changing the
exponent of the SFR power law ($n$) affects the solutions only slightly;
$n$ just determines the time scale for which a certain metallicity is
reached.  Lowering the efficiency parameter $\nu$ shifts the solutions
to slightly lower metallicities.  Similarly, steepening the IMF shifts
the curves to lower metallicities and larger C/O ratios since fewer
massive stars are produced.

By using the properties of galaxies at the current epoch, we can
constrain many of the free parameters in the models.  To be consistent
with the net yields derived for the Galaxy, we choose models with $0.011
\leq y_z \leq 0.036$.  For the closed box models, the implied net yields
are $y_z=2.72(Z\mu)_{max}$, while for the infall models
$y_z=3.15(Z\mu)_{max}$.  As a second constraint, we use the relationship
found by Garnett \etal (1995): $\log(\ac/\io) \propto B\log(\io/{\rm
  H})$, where $B=0.43\pm0.09$ over the metallicity range of $-4.64 \leq
\log(\io/{\rm H}) \leq -3.6$ ratio.  An additional constraint is the
value of the O/C ratio at the lowest metallicity data point of Garnett
\etal (1995): $6\leq \io/\ac \leq 13$ at $\log(\io/{\rm H})=-4.64$.
Using the metallicities and gas fractions of typical massive spiral
galaxies in the local universe, we adopt $1 \leq Z/Z_{\odot} \leq 5$ and
$0.01 \leq \mu \leq 0.1$ as the fourth and fifth constraints.  In
Table~2 we list the computed values of these 5 constraints for each of
the models.  Only the infall models of M92 (Run 10, 11, \& 12) and the
closed box model with $\alpha=0$ (Run~5) satisfy all 5 of the
constraints.  Current stellar evolutionary theories argue for values of
$\alpha \ga 1.5$ (\eg Chieffi \etal 1995), so the $\alpha=0$ models are
most likely not realistic and are included in this study mainly for
comparison purposes.  Hence, these results indicate that infall models
using the stellar yields of M92 best fit the observed properties of the
Milky Way.  These results are consistent with the models of Prantzos
\etal (1994) and Carigi (1994) which also addressed Galactic chemical
evolution using the yields of M92.  Unlike the models based on the
yields of M92, the models based on the yields of WW95 do not predict the
observed evolution of the C/O ratio with metallicity found for {\sc Hii}
galaxies.  These results underscore the importance of stellar winds,
which are neglected in the WW95 yields, on the specific issue of the
evolution of the C/O ratio.  Although the WW95 models are not well
tailored to understanding the evolution of the C/O ratio, these models
do equally as well as the M92 models in describing the evolution of the
metallicity and the gas fraction.  In comparison with previous work, we
find that the model parameters of Run~24, which are similar to the values
used by Timmes et al. (1995), yield consistent results.

In Table~3 we list the time it takes for the models to achieve
interesting stages of evolution.  We include the age at which one half
of the final total baryonic mass is in stars and stellar remnants
($\tau^{*}_{1/2}$), the age at which $ZM_{g}$ is a maximum, and the age
at which the system reaches solar metallicity.  Generally, solar
metallicity is achieved after $\tau^{*}_{1/2}$.  However, for the infall
models using the yields of M92, solar metallicities are achieved before
$\tau^{*}_{1/2}$ and before the peak of gaseous metals.  In summary, all
three of these epochs occur at a similar time for each of the models and
occur roughly within the first 2~Gyr.  Granted, not all galaxies evolve
this quickly.  Dwarf galaxies (Hunter \& Gallagher 1985) and low surface
brightness galaxies, such as Malin 1 (Impey \& Bothun 1989), certainly
do not.  The models given in this paper are more representative of
massive galaxies which undergo prodigious star formation at early times.

The current observations and the results of these modeling efforts are
certainly not sufficient to determine a unique solution governing
chemical evolution.  However, this is not the goal of our work.  The
principal motivation behind these studies has been to determine the
general evolution of carbon, oxygen, and metallicity and how this
evolution is affected by varying input parameters.  The basic results
are summarized below.  The models are consistent with the observations
and predict that oxygen should be much more abundant than carbon at
early epochs (O/C$\ga 10$).  The evolution of carbon lags behind that of
oxygen since it is produced at later times, from intermediate--low mass
stars and from the winds of metal--rich massive stars.  In addition to a
different composition, the ISM of young galactic systems should contain
a mass of gaseous metals which is greater than that found at the current
epoch where the gas has been depleted due to star formation.  Using
these basic results, we now shift our attention to the molecular gas
content of young galactic systems.

\section{Molecular Chemistry}

The models calculating the gas--phase abundances within dark
molecular clouds consist of a sophisticated network of order one
thousand time--dependent chemical reactions (Prasad \& Huntress 1980;
Graedel, Langer, \& Frerking 1982; Herbst \& Leung 1986 [HL86]; Herbst
\& Leung [HL89]; LG89; MH90).  These models are sensitive to a variety
of parameters such as the density, temperature, ionization field, and
the initial chemical composition.  However, the dominate parameter for
determining the relative abundances of the carbon and oxygen species is
the gas--phase carbon to oxygen ratio, (C/O)$_{g}$ (LG89).  In
Figure~\ref{lg89fit}, we show the computed abundances of the C, O,
O$_2$, and CO species as a function of the C/O ratio at
approximately solar metallicity for the steady--state solutions of LG89.
This figure demonstrates several important tendencies.  The O$_2$/H$_2$
ratio is extremely sensitive to the (C/O)$_g$ ratio, while the CO/H$_2$
ratio is relatively constant.  For high (C/O)$_g$ ratios, there is
virtually no molecular oxygen since most of the free oxygen combines
with carbon to form CO.  For (C/O)$_g \la 0.5$, there is an excess of
oxygen after the formation of CO which may combine to produce O$_2$.

\subsection{The Abundance of O$_2$ in the Local Universe}
\label{sec-localo2}

Molecular oxygen has yet to be detected conclusively outside the solar
system despite the efforts of several observers.  Due to atmospheric
attenuation, Galactic studies have been limited to searches of the rarer
isotope $^{16}$O$^{18}$O, which have provided upper limits of O$_2$/CO
$\la 0.1$ (Liszt \& Vanden Bout 1985 [LV85]; Goldsmith \etal 1985;
Combes \etal 1991 [C91]; Fuente \etal 1993; Mar\'{e}chal et al. 1997).
Extragalactic searches for the redshifted 118.75 GHz $^{16}$O$_2$
emission line ($z \ga 0.03$) have provided more sensitive limits, but
they also have been unsuccessful (Liszt 1985, Goldsmith \& Young 1989
[GY89], C91, Liszt 1992 [L92]).  Recently, Combes \& Wiklind (1995) have
reported a very sensitive nondetection of O$_2$ ($N(\oo)/N(\co) <
0.014$) from an absorption line experiment toward a background
mm--wavelength continuum source.  The current most stringent
observational limit is $N(\oo)/N(\co) < 0.01$ for the $z=0.026$ galaxy
NGC 6240 (C91).

If the gas--phase C/O ratio is similar to the solar ratio of
C/O$\simeq 0.4$, then the observed upper limits of the O$_2$/CO ratio
are less than the values predicted by the steady--state theoretical
models ($\oo/\co\sim 0.5$).  Several ideas have been put forth to
explain this inconsistency.  These ideas can be divided into two
broad classes: (1) deviations of gas--phase abundances from solar; and
(2) limitations of the gas--phase steady--state chemistry models.

Perhaps the solar C/O ratio is not typical within the molecular gas
clouds previously searched for O$_2$.  If the gas--phase C/O ratio is
increased by only a factor of 2, all of the observational upper limits
of O$_2$ are consistent with the steady--state expectations.  For
example, the average value of the C/O ratio determined by several
measurements within the Orion Nebula is C/O $=0.71\pm0.18$ (Meyer \etal
1994 (M94) and references therein).  The solar C/O ratio is also smaller
than the value of C/O $\sim 0.6$ predicted from extrapolating the
relation found by Garnett \etal (1995).  One of the difficulties in
directly comparing the C/O ratio and the O$_2$/CO limits is that the
elemental abundances are determined by absorption lines through the
diffuse ISM toward stars, stellar abundances, or from the studies of
{\sc Hii} regions.  The abundances within the dense molecular regions
may be significantly different.  In any case, the C/O ratio is not
constant within the ISM, and typical measurements range between C/O
$\sim 0.4 - 0.8$ (e.g., Cardelli \etal (1993) [C93]).  These differences
may arise due to local environmental conditions.  For example, the Sun
may have formed in the late stages of the life of its parent molecular
cloud, and the solar abundances may have been enriched from earlier
supernovae within the cloud (Cunha \& Lambert 1994).  Since supernova
ejecta from massive stars are more abundant in oxygen than in carbon
(\S~\ref{sec-yields}), the solar C/O ratio may be lower than the typical
ISM abundance ratio.

The situation is even less clear for extragalactic measurements where
the physical conditions and chemical properties could vastly differ from
Galactic clouds.  The extragalactic searches for O$_2$ have concentrated
on the bright cores of CO--rich star--forming galaxies.  These central
regions of the galaxies are expected to have at least roughly solar
metallicity and could possibly have metallicities as large as
$5Z_{\odot}$ (\eg Pagel \& Edmunds 1981).  Since C/O varies roughly as
$Z^{1/2}$ (\S\ref{sec-chem_results}), the carbon to oxygen ratio is
expected to be larger than solar.  For the galaxy NGC~7674, Kraemer
\etal (1994) [K94] find that the best model fit to their IUE spectra
suggests that C/O$\sim 3/2$.  Therefore, it is not surprising that O$_2$
was not observed in this source by L92.  If similarly large C/O ratios
are found for the center of other galaxies, then the observational
deficiency of extragalactic O$_2$ in the local universe would not
contradict the theoretical estimates.

An additional method suggested for increasing the gas--phase C/O ratio
involves the selective depletion of oxygen (C91).  Several mechanisms
have been proposed for excess oxygen depletion.  The gas phase oxygen
abundance may be depleted by large quantities of solid H$_2$O, CO$_2$,
CO, O$_2$, and silicates on the mantles of grains (d'Hendecourt \&
Jourdain de Muizon 1989; Ehrenfreund et al. 1993; de Graauw et al. 1996;
Whittet et al. 1996).  Since both CO and O$_2$ have similar binding
energies to the dust grains, these species are expected to have similar
levels of depletion onto grain surfaces (Bergin, Langer, \& Goldsmith
(1995) [BLG95]).  Barring any additional effects, the gaseous O$_2$/CO
ratio would remain roughly constant as depletion increases.  Blake \etal
(1987) suggests that even if C and O are adsorbed at equal rates onto
grains to form primarily CH$_4$ and H$_2$O, respectively, the nonpolar
CH$_4$ molecule will evaporate more easily than the polar H$_2$O
molecule, resulting in larger gas--phase C/O ratios.  However,
observations are inconsistent with large H$_2$O abundances.  Only 10\%
of the oxygen is contained within solid H$_2$O in dense clouds (Smith,
Sellgren, \& Brooke 1993), and even less is contained within gaseous
H$_2$O.  Using observations with the Kuiper Airborne Observatory,
Zmuidzinas \etal (1996) conclude that much less than one percent of the
oxygen is in gaseous H$_2$O within the cool quiescent molecular
envelopes.  In hot molecular cores, only a few percent of the oxygen is
in gaseous H$_2$O (Zmuidzinas et al. 1996; Gensheimer, Mauersberger, \&
Wilson 1996; van Dishoeck \& Helmich 1996), which is consistent with the
evaporation of solid H$_2$O from the grain mantles or may suggest the
importance of high temperature gas--phase chemistry (van Dishoeck \&
Helmich 1996).  In any case, current observations do not support the
scenario of excess oxygen depletion as a mechanism for significantly
increasing the C/O ratio with respect to the solar value.

If the solar C/O ratio is the typical gas--phase value within molecular
clouds, then the most likely reason for the observational deficiency of
O$_2$ involves limitations in the gas--phase steady--state chemistry
models.  More recent models have included grain surface adsorption and
desorption (BLG95) and have included reactions on the surfaces of grains
(Hasegawa \& Herbst 1993; Willacy \& Williams 1993).  Unfortunately, the
steady--state solutions of these models still have difficulties
explaining the observed molecular abundances.  In particular, the models
that include networks of reactions on grain surfaces vastly under
produce the gas--phase molecular abundances in dense clouds.
Observations of the dense Taurus region have shown that only 5\% to 40\%
of the total number of CO molecules exist on grains (Whittet 1993) with
a gas phase abundance ratio of $\co/\hh \sim 10^{-4}$ (Frerking, Langer,
Wilson 1982), whereas surface chemistry models predict much lower
gas--phase abundances (e.g., $\co/\hh \sim 10^{-9}$; Hasegawa \& Herbst
1993).  Models that neglect surface reactions, but include adsorption
and desorption, do a much better job of yielding sufficient quantities
of gas--phase CO (BLG95).  However, the models of BLG95 predict similar
levels of depletion for CO and O$_2$, and the resulting O$_2$/CO ratio
does not vary from the gas--phase chemistry models of LG89 and MH90
(Fig.~\ref{lgmh}).  Therefore, grain chemistry models, by themselves,
cannot currently explain the apparent deficiency of O$_2$ in Galactic
clouds.

An alternative explanation for the observed deficiency of O$_2$ may
involve turbulent diffusion and mixing (Xie, Allen, \& Langer 1995
[XAL95]; Chi\`{e}ze \& Pineau des For\^{e}ts 1989; Pineau des
For\^{e}ts, Flower, \& Chi\`{e}ze 1992) [PFC92]).  Mixing of dense
interior gas with the outer layers prevents the clouds from reaching
their steady--state solutions ($t>10^{7}\yr$), resulting in solutions
similar to those found at earlier times $(t\sim 10^{5} \yr$; \eg MH90;
HL89).  For these models the atomic carbon abundances are much larger,
and the O$_2$ abundances can be lowered by several orders of magnitude.
The O$_2$ abundance is extremely sensitive to the abundance of atomic
carbon since the O$_2$ molecules are destroyed by atomic carbon to form
CO.  The carbon atoms also compete with the oxygen atoms in reacting
with OH molecules, which decreases the rate of O$_2$ formation.  The
observations of the atomic carbon lines in M82 (Schilke et al. 1993;
Stutzki et al. 1997) and NGC 253 (Harrison et al. 1995) suggest enhanced
atomic carbon abundances in these nearby starburst systems.  The high
atomic carbon abundances may result from a turbulent, clumpy medium where
the clouds are repeatedly exposed to UV radiation (Stutzki et al. 1997),
possibly suggesting the applicability of mixing scenarios and the
early--time solutions.  The situation could be vastly different for
chemically young galaxies where the O/C$\gg 1$.  For O/C$\ga 10$, the
total carbon abundance is lowered sufficiently such that the early--time
solutions are expected to be similar to the steady--state solutions
(Fig.~\ref{lgmh}).

In Figure~\ref{lgmh} we show the O$_2$/CO ratio as a function of the O/C
ratio computed by many theoretical models as well as the current
observational limits.  For the LG89 and MH90 models, the gas--phase
steady--state solutions are for molecular clouds with large values of
extinction ($A_{V}\sim10$).  The dark cloud solutions of LG89 and MH90
are fitted by the following function:
\begin{equation}
\label{eq:lgmhfit}
\log\left(\frac{\oo}{\co}\right) = -0.357 +
0.957\log\left(\frac{\io}{\ac}\right) -1.426\left(\frac{\ac}{\io}\right)^{2}.
\end{equation}
The coefficients are determined by a singular value decomposition
least--squares fit to the data (Press \etal 1986).  In the limit as
$\io/\ac \rightarrow \infty$, equation (\ref{eq:lgmhfit}) conserves the
number of total oxygen atoms and has the correct dependency on the O/C
ratio [$\oo/\co \la 1/2(\io/\ac)$; LV85].  Equation (\ref{eq:lgmhfit})
fits remarkably well to the data of LG89 and MH90 (see Fig.~\ref{lgmh}),
even though the LG89 solutions are for Galactic clouds, while the
solutions of MH90 are for clouds in the LMC and SMC which have lower
input metallicities and higher UV radiation fields.  From these results
we deduce that the O$_2$/CO ratio is primarily determined by the O/C
ratio and does not depend significantly on the metallicity within dark
clouds.  All of the other theoretical points in Figure~\ref{lgmh} are
normalized to the data of LG89 and MH90 using their respective
steady--state solutions.  The fit displayed for the early--time
solutions is made using only the data of MH90 and has the same
functional form as equation (\ref{eq:lgmhfit}).  The uncertainties in
the O/C ratio for NGC~6240 and the galaxies of L92 and GY89 are
estimated by assuming that these galaxies have a central metallicity of
$Z/Z_{\odot}=2\pm\stackrel{3}{_{1.5}}$ and that $Z\propto (\ac/\io)^2$
(\S\ref{sec-chem_results}).  For the molecular regions of $\rho$ Oph, we
adopt the O/C ratio derived toward $\zeta$ Oph as a ``typical'' ISM
value (C93), and this ratio may not be applicable for $\rho$~Oph.

\subsection{The Evolution of the O$_2$/CO Ratio}
\label{sec-o2corat}

Even though molecular oxygen has yet to be observed in the local
universe, O$_2$ may be more abundant at early cosmological epochs.  At
early times when O/C$\gg 1$, large quantities of O$_2$ could be produced
simply because of an overabundance of oxygen relative to all of its
antagonist species.  In Figures~\ref{o2cotall} \&~\ref{o2cozall} we plot
the expected O$_2$/CO ratios in dark clouds as a function of the age and
metallicity for the steady--state chemistry models.  In general, the
evolution of the O$_2$/CO ratio is most sensitive to the IMF and star
formation efficiency and is not as dependent on $n$ and $\tau_f$.
Basically, all models indicate that $\oo/\co \ga 3$ within dark
molecular clouds at early epochs.  The models based on the stellar
yields of M92 predict a rapid decrease of the O$_2$/CO ratio when
$Z\rightarrow Z_{\odot}$ due to the excess carbon production from the
winds of massive stars.  The models based on the stellar yields of WW95
predict a slower decrease of the O$_2$/CO ratio since these models
neglect the contribution of carbon from the winds of massive stars.  The
model solutions in Figures~\ref{o2cotall} \&~\ref{o2cozall} are only
applicable for molecular regions which are sufficiently shielded.  In
order to estimate the global molecular abundances in galaxies associated
with an ensemble of molecular clouds with various values of total
extinction, we must correct for the effects of photodissociation.

\subsubsection{Correction for Photodissociation}

The rate of photodissociation depends on the ambient UV field, and hence
the SFR, and on the amount of metals shielding the molecular gas.  Since
the photodissociation rates for O$_2$ are larger than the rates for CO
(van Dishoeck 1988), CO could exist in diffuse molecular clouds, while
O$_2$ would not.  Therefore, the spatial extent of the O$_2$ regions are
expected to be smaller than the CO emission regions.  Similarly, the
extent of the CO regions is expected to be smaller than that of the
H$_2$ regions since the H$_2$ molecules are more strongly self--shielded
(van Dishoeck \& Black 1988; Maloney \& Black 1988).  Assuming spherical
geometry, we divide the molecular clouds into 3 distinct regions: (1)
the central regions containing O$_2$, CO, and H$_2$, (2) the
intermediate regions containing CO and H$_2$, and (3) the outer regions
which contain only H$_2$.  The relative extent of these regions depends
on the radiation field and metallicity (i.e., the amount of dust).
Maloney \& Black (1988) have shown that near the edge of the CO regions,
the amount of carbon contained within CO is strongly nonlinear with
metallicity.  For an average Galactic UV field (\eg Draine 1978), we
expect these edge effects to be important for cloud depths of $0 < A_{V}
< 1.5$ (van Dishoeck \& Black 1988).  In deeper regions with extinctions
larger than this critical value of $A_{\co} = 1.5$, the gas is
sufficiently shielded so that virtually all of the carbon is contained
within CO.  By using the results of BLG95, we adopt a critical
extinction value for O$_2$ of $A_{\oo} = 5$.  Therefore, the
molecular chemistry solutions of the O$_2$/CO ratios previously
presented as Figures~\ref{o2cotall} \&~\ref{o2cozall} are only valid for
dark clouds with $A_V \ga 5$.

The critical extinction boundaries of the CO and O$_2$ molecular regions
depend on the ambient UV field.  The aforementioned values of
$A_{\co}=1.5$ and $A_{\oo}=5$ are for an average Galactic UV field.  In
regions with more intense UV fields, the critical extinction values will
be significantly larger.  Using the study of photodissociation regions
by Hollenbach, Takahshi, \& Tielens (1991), we estimate that $A_{\co}
\propto G_{o}^{0.1}$, where $G_{o}$ is the UV field relative to the
solar neighborhood.  Assuming that $G_{o}$ is proportional to the SFR
and that a similar relationship exists for both the CO and O$_2$
molecules, we expect that the critical extinction values for CO and
O$_2$ should vary roughly as
\begin{equation}
\label{eq:avco}
A_{\co} = 1.5 \left(\frac{SFR}{5 \msun \yr^{-1}}\right)^{0.1},
\end{equation}
and 
\begin{equation}
\label{eq:avoo}
A_{\oo} = 5.0 \left(\frac{SFR}{5 \msun \yr^{-1}}\right)^{0.1}.
\end{equation}
In equations (\ref{eq:avco}\&\ref{eq:avoo}) we use a Galactic SFR
of $5\msun \yr^{-1}$ to normalize the critical extinction values for the
average Galactic UV field.  

In order to correct the global O$_2$/CO ratio for these
photodissociation effects, we must estimate the mass fraction of the
molecular gas contained within the O$_2$ and CO emission regions.  Most
of the molecular mass of galaxies is contained within giant molecular
clouds (GMCs).  We adopt a GMC mass distribution function of $dN/dM
\propto M^{-3/2}$ with a lower mass limit of $10^{4}\msun$ and an upper
limit of $2\times 10^{6}\msun$ (Solomon \& Rivolo 1987).  Assuming a
uniform density and using the size density relationship found for GMCs
by Sanders, Scoville, \& Solomon (1985), we find that the total
extinction for a cloud at solar metallicity $(A_{\odot})$ and with a
mass $M$ in solar units is
\begin{equation}
\label{eq:ami}
A_{\odot} = 15(M /5.8\times 10^{5}\msun )^{1/9},
\end{equation}  
where a $5.8\times 10^{5} \msun$ GMC (including a factor of 1.36 for
helium) has a total H$_2$ column density of $N(H_2) = 1.5 \times 10^{22}
\cm^{-2}$.

Assuming spherical geometry and integrating over an entire GMC of mass $M$,
the total volume--weighted fraction of the H$_2$ mass coexisting with
molecule $x$ is
\begin{equation}
\label{eq:fmxhh}
\frac{M(\hh)^{x}}{M(\hh)}  = 3/R^{3} \int_{0}^{R} \chi_{x}(r) r^2dr,
\end{equation}
where $\chi_{x}$ is the abundance of species $x$ relative to its dark
cloud abundance and $R$ is the radius of the molecular cloud.  Upon a
change of variable $\tau/A_{total} = 1 - r/R$, the depth--dependence can
be calculated as a function of the extinction $\tau$.  The total
extinction of a cloud ($A_{total}$) varies linearly with the average
density of dust particles (Spitzer 1978), which is expected to vary
linearly with metallicity (Issa, MacLaren, \& Wolfendale 1990; van den
Hoek \& de Jong 1992).  We adopt $A_{total}=(Z/Z_{\odot})A_{\odot}$,
where $A_{\odot}$ is calculated from equation (\ref{eq:ami}).  Using
the study of Maloney \& Black (1988) to determine the CO abundance for low
extinction regions near the edges of molecular clouds, we estimate that
\begin{equation}
\label{eq:edge}
\chi_{x} = \left\{  \begin{array}{ll} (\tau/A_{x})^{3} & \tau <
    A_{x}\\ 1 & \tau \geq A_{x}, \end{array}
\right.
\end{equation}
where $A_{x}$ represent the critical extinction values of $A_{\co}$ and
$A_{\oo}$ given by equations (\ref{eq:avco} \&\ref{eq:avoo}).  For
metallicities less than $(A_{\co}/A_{\odot})Z_{\odot}$, the total
extinction of a GMC will be less than the critical extinction value for
CO.  In this case, the fractional mass of the CO regions is determined
by the photodissociation edge--effects.  Likewise, for metallicities
less than $(A_{\oo}/A_{\odot})Z_{\odot}$, the total extinction of the
GMCs will be less than the critical extinction value for O$_2$, and
the mass of the O$_2$ regions is dominated by the edge--effects.  The
total correction for photodissociation is calculated by summing over the
GMC distribution function:
\begin{equation}
\label{eq:fmh2co}
\frac{M(\hh)^{\co}}{M(\hh)} = \sum_{i=1}^{n} f_i
\left(\frac{M(\hh)^{\co}}{M(\hh)}\right)_{i},
\end{equation}
and
\begin{equation}
\label{eq:fmh2o2}
\frac{M(\hh)^{\oo}}{M(\hh)} = \sum_{i=1}^{n} f_i
\left(\frac{M(\hh)^{\oo}}{M(\hh)}\right)_{i},
\end{equation}
where $f_i$ are the weighting coefficients normalized to unity.  The
$M(\hh)^{\co}/M(\hh)$ ratio is the fraction of the H$_2$ mass coexisting
with CO, and $M(\hh)^{\oo}/M(\hh)$ is the fraction of the H$_2$ mass
coexisting with O$_2$.

\subsubsection{Other Scenarios Besides Steady State}

Given the possible limitations of the steady--state solutions (\S
\ref{sec-localo2}), we analyze 2 additional scenarios for the evolution
of the O$_2$/CO ratio: (1) the early--time scenario where the clouds do
not reach their steady--state solutions and (2) the selective depletion of
oxygen model.  The early--time
model solutions are presented in Figure~\ref{lgmh}.  For O/C$\leq 10.1$,
the formal fit of the early--time scenario is given by
\begin{equation}
\label{eq:etfit}
\log\left(\frac{\oo}{\co}\right) = -0.734 +
1.378\log\left(\frac{\io}{\ac}\right) -6.740\left(\frac{\ac}{\io}\right)^{2}.
\end{equation}
For $\io/\ac > 10.1$, we assume that the early--time solutions are
equivalent to those of steady state (eq.~[\ref{eq:lgmhfit}]). 

In the scenario of selective oxygen depletion, the gas--phase
(C/O)$_{g}$ ratio is enhanced relative to the total elemental C/O ratio.
Since the proposed enhancement mechanisms of (C/O)$_{g}$ are primarily
associated with grain processes, these effects are not expected to be as
important at early cosmological epochs when the metallicity and dust
abundances are lower on average.  Studies of the nearby galaxies
indicate that the dust--to--gas ratio is approximately proportional to
the average metallicity (Issa et al. 1990; van den
Hoek \& de Jong 1992).  Assuming a linear correlation with metallicity, we
let $\gamma$ be a free parameter describing the enhancement of C/O
ratio, so that
\begin{equation}
\label{eq:cosod}
(\ac/\io)_{g}=[\gamma(Z/Z_{\odot}) + 1](\ac/\io).
\end{equation}
For the solar abundance ratio of C/O$\simeq 0.4$, the observational
limit for the O$_2$/CO ratio found by C91 would suggest that $\gamma \ga
2$.  We adopt this limiting case ($\gamma=2$) for the selective
depletion of oxygen scenario.  The dark cloud O$_2$/CO ratio in this model is
determined by substituting the (C/O)$_g$ ratio given by equation
(\ref{eq:cosod}) into equation (\ref{eq:lgmhfit}).

For the three scenarios of (a) steady state, (b) early times, and (c)
selective oxygen depletion, we use equations (\ref{eq:fmh2co}
\&\ref{eq:fmh2o2}) to determine the global volume--weighted O$_2$/CO abundance
ratio corrected for photodissociation effects as a function of
metallicity and the SFR.  The global abundances are calculated using
\begin{equation}
\label{eq:global}
\left(\frac{\oo}{\co}\right)_{global}=
\left(\frac{M(\hh)^{\oo}}{M(\hh)^{\co}}\right)
\left(\frac{\oo}{\co}\right)_{dark},
\end{equation}
where $(\oo/\co)_{dark}$ is the O$_2$/CO abundance ratio computed for
the dark cloud solutions.  In Figure~\ref{ch_mol4} we show the evolution
of the O$_2$/CO ratio for these scenarios compared with the dark cloud
steady--state solutions.  We choose 4 different models (Run~1, 10, 13,
\& 22) to represent examples of the closed box and infall models for
both the M92 and WW95 stellar yields.  The volume--weighted correction
for photodissociation reduces the global O$_2$/CO ratios by a factor of
approximately 30 at low metallicities.  These corrections are fairly
uncertain for low metallicities and greatly depend on the adopted
parameterization describing the edge--effects (eq.~[\ref{eq:edge}]) and
on the structure of the clouds.  At solar metallicities using the
stellar yields of WW95, only scenario~(c) is consistent with the
Galactic observational upper limit of the O$_2$/CO ratio.  For the M92
models, all 3 scenarios are roughly consistent with the observational
limits due to the higher carbon yields from stellar winds at solar
metallicity.  Interestingly, both the scenarios (b) \& (c) for the M92
models predict a similar evolution of the O$_2$/CO ratio.

\subsection{The Evolution of the CO, O$_2$, and Dust Abundances}
\label{sec-coo2dust}

The curves presented in Figures~\ref{o2cotall}, ~\ref{o2cozall} \&
~\ref{ch_mol4} show the evolution of the $\oo/\co$ ratio, but they do
not indicate how much molecular gas is present.  In order to determine
the individual CO and O$_2$ abundances, we must first estimate several
important parameters which are currently not known very well as a
function of evolution and metallicity.  These parameters include the
molecular to neutral gas fraction [$M(\hh)/M(\hi)$] and the CO abundance
relative to H$_2$.  The determination of these factors is complicated by
uncertainties in the empirical relation used to determine $M(\hh)$ from
the observed CO luminosity.  The first step in analyzing these
complications and estimating the individual abundances of CO and O$_2$
is to estimate the molecular gas fraction.

The molecular gas fraction is uncertain for young galactic systems. The
recent CO detections at high redshift are promising and indicate that at
least some young galactic systems contain significant quantities of
molecular gas (Brown \& Vanden Bout 1991; Solomon \etal (1992);
Barvainis \etal 1994a; Ohta et al. 1996; Omont et al. 1996b).  Although
optical absorption line studies along absorbing columns through damped
\lya systems suggest low molecular hydrogen abundances (Levshakov et al.
1992 and references therein; Ge \& Bechtold 1997), these optical
absorption line studies are strongly biased against detecting QSOs
behind large columns of dust (Fall \& Pei 1995; Boiss\'{e} 1995).
Therefore, the global molecular masses associated with these young
systems may be much larger than what would be inferred by only the
optical observations.  The global H$_2$/H{\sc i} mass ratio is an
important parameter for the models presented in this paper.  Radio
wavelength CO and H{\sc i} emission line studies, in principle, can
directly determine this mass ratio.  Unfortunately, the 21--cm H{\sc i}
emission line, from which the H{\sc i} mass is derived, is generally too
weak to be observed at high redshift, and these observations are also
often plagued by radio frequency interference.  Since the global
H$_2$/H{\sc i} mass ratio has yet to be measured for any high redshift
system, we apply the results determined from the studies of nearby
galaxies.

In a compilation of low redshift CO data, Young \& Scoville (1991) find
that galaxies with early morphological types (S0--Sa) have molecular to
neutral hydrogen mass ratios of $M(\hh)/M(\hi)= 4.0\pm1.9$, whereas
late--type galaxies (Sd--Sm) have ratios of $M(\hh)/M(\hi)= 0.2\pm0.1$.
Whether this result is a true representation of the molecular gas
content as a function of morphological type is unclear.  It is possible
that these results reflect differences in the CO to H$_2$ conversion
factor (Roberts \& Haynes 1994).  Young \& Scoville (1991) assume a
constant $M(\hh)/L(\co)$ ratio determined empirically from the study of
Galactic clouds.  This relation may underestimate the true amount of
molecular gas in late--type galaxies which have lower average
metallicities.
 
Recent observations indicate that the Galactic $M(\hh)$ to $L(\co)$
conversion factor does underestimate the molecular gas mass in
metal--poor dwarf galaxies (Wilson 1995; Arimoto \etal 1996).  The
metallicity dependency implied by these observations suggests that
$M(\hh)/L(\co) \propto (Z/Z_{\odot})^{-0.7}$.  This relationship is
calibrated assuming that the total virial mass of $M(\hh)$ is given by
the width of the CO line.  Since the CO observations only measure the
mass within the CO emission regions [$M(\hh)^{\co}$], not the
potentially large envelopes of H$_2$ which contain no CO, we divide the
molecular hydrogen gas into regions which contain CO and regions of
lower extinction which do not have CO (\S\ref{sec-o2corat}).  Assuming
that 25\% of the total gaseous mass is He, we find that
\begin{equation}
\label{eq:mgall}
0.75 M_{g}= M(\hh)^{*} + M(\hh)^{\co} + M(\hi),
\end{equation}
where $M(\hh)^{*}$ is the mass of H$_2$ in the regions without CO and
$M(\hh)^{\co}$ is the mass of H$_2$ coexisting with the CO.  We neglect
the dust mass in equation (\ref{eq:mgall}) since the gas--to--dust mass
ratios are typically of order 100 or larger.  Solving for the total mass
of CO molecules, we find that
\begin{equation}
\label{eq:mco}
M_{\co}(t)=10.5[M(\hh)/M(\hh)^{\co} + M(\hi)/M(\hh)^{\co}]^{-1} M_g(t)
\xi_{\co}(t), 
\end{equation}
where $\xi_{\co}$ is abundance of CO relative to H$_2$ within the CO
emission regions.  The $M(\hh)/M(\hh)^{\co}$ ratio is given by equation
(\ref{eq:fmh2co}).  By using the CO data for galaxies of different
morphologies (Young \& Knezek 1989) and the average metallicities as a
function of morphological type (Roberts \& Haynes 1994), we find an
approximate relationship of $M(\hh)^{\co}/M(\hi) \sim (Z/Z_{\odot})$
assuming the Galactic $L(\co)$ to $M(\hh)^{\co}$ conversion factor.
After correcting the conversion factor using the metallicity
relationship found by Wilson (1995) and Arimoto \etal (1996), we find
that $M(\hh)^{\co}/M(\hi)$ is approximately constant;
$M(\hh)^{\co}/M(\hi) \sim 2$.  We adopt this value for equation
(\ref{eq:mco}).

The models of MH90 suggest that the CO abundance relative to H$_2$
within the CO emission regions scales directly with the elemental carbon
abundance and is independent of the oxygen abundance as long as O/C$ >
1$.  For O/C$ <1$, most of the oxygen is tied up in CO, and $\xi_{\co}$
is expected to scale with the oxygen abundance.  By using the results of
the MH90 models and normalizing the data to the Galactic CO abundance of
$\xi_{\co}=10^{-4}$ (\eg Young 1990), we adopt
\begin{equation}
\label{eq:xico}
\xi_{\co} = \left\{ \begin{array}{ll}
(Xc/Xc_{\odot})10^{-4} & \io/\ac \geq 1 \\  
(Xo/Xo_{(o/c=1)})(Xc_{(o/c=1)}/Xc_{\odot})10^{-4} & \io/\ac < 1,\end{array}
\right.
\end{equation}
where $Xc$ and $Xo$ are the carbon and oxygen mass fractions calculated
in the models of \S2.  The solar carbon mass fraction is $Xc_{\odot} =
0.00303$ (Grevesse \& Anders 1989), and $Xc_{(o/c=1)}$ and
$Xo_{(o/c=1)}$ are the carbon and oxygen mass fractions when O/C$=1$.

After calculating the CO mass using equation (\ref{eq:mco}), we compute
the mass in molecular oxygen using
\begin{equation}
\label{eq:mo2}
M_{\oo}(t)=\frac{8}{7}\left[ \frac{\oo}{\co}(t)\right] M_{\co}(t),
\end{equation}
where $\oo/\co$ is derived for the different scenarios described in
\S\ref{sec-o2corat}.

We calculate the dust mass in a straightforward fashion.  We assume that
the dust--to--gas ratio varies linearly with the average metallicity
(Issa et al. 1990; van den Hoek \& de Jong 1992), and
we adopt the Galactic gas--to--dust ratio of $M_{g}/M_{dust} \sim 100$
(Savage \& Mathis 1979; Mathis 1990; Tielens et al. 1996) to normalize
the relationship;
\begin{equation}
\label{eq:mdust}
M_{dust}(t) = 0.01 (Z(t)/Z_{\odot}) M_{g}(t),
\end{equation}
where $M_{g}/M_{dust} = 100$ for solar metallicity.  Uncertainties in
the derived gas--to--dust ratios depend significantly on the composition
of the grains which can have different frequency--dependent absorption
coefficients that depend on temperature (e.g., Agladze et al. 1996).
For the nearby IRAS galaxies, the derived gas to warm dust ratios are
about a factor of 10 larger than the Galactic value, suggesting that
most of the dust in galaxies is colder than 30 K (Devereux \& Young
1990; Sanders, Scoville, \& Soifer 1991).  Even though more research
into the properties of dust grains and sensitive observations at
$\lambda > 100 \micron$ are required to constrain the amount of cold
dust in galaxies, equation (\ref{eq:mdust}) is currently an adequate
approximation.

In Figures~\ref{ch_z4}\&\ref{ch_mu4}, we plot the evolution of the mass
of CO, O$_2$, and dust as a function of metallicity and the gas
fraction.  The molecular masses fall off rapidly with decreasing
metallicity due to the lack of shielding, while the dust masses decrease
linearly with decreasing metallicity (Fig.~\ref{ch_z4}) . We expect the
largest O$_2$ masses in the metallicity range of $0.1 < Z/Z_{\odot} <
1$.  Within this range of metallicities, there is enough dust to provide
adequate shielding of the O$_2$ molecules, but the metallicity is low
enough for high O/C ratios.  For the models consistent with Galactic
observations, the maximum of the O$_2$ mass occurs at early epochs ($\mu
\ga 0.5$, $z>2$).  The maximum dust mass occurs later ($\mu \sim 0.4$)
and is followed in time by the maximum CO mass (Fig.~\ref{ch_mu4}).
This evolutionary progression is expected since the production of CO
lags slightly behind the dust due to the increase of shielding with
metallicity, while the O$_2$ abundances are largest at the lower
metallicities when O/C$\gg1$.  By comparing the different models, we
find several general results. Since the carbon stellar yields are
smaller in the WW95 models, the $M(\oo)/M(\co)$ ratios for the WW95
models are larger than those for the M92 models.  The dust masses in the
M92 models are larger than the dust masses in the WW95 models, and these
discrepancies are consistent with the differences between their
effective net yields (Table~3).  The situation is more subtle for the CO
mass.  As with $M(dust)$, the CO masses are larger in the M92 models.
However, because the maximum $M(\co)$ value occurs at a larger
metallicity in the M92 models, the ratios of the $M(\co)$ values (eq.
[\ref{eq:mco} \& \ref{eq:xico}]) between the M92 and WW95 models are
larger than the ratios of their effective net yields.

For the Galactic disk, which has a total mass of $6\times 10^{10}\msun$
(Bahcall \& Soneira 1980) and an approximate H$_2$ mass of $M(\hh)^{\co}
\sim 2 \times 10^{9}$ (Solomon \& Rivolo 1987), we estimate
$M_{\co}/M_{T} \sim 10^{-4}$ and $M_{dust}/M_{T}\sim 10^{-3}$ assuming
$\mu=0.1$ (Rana \& Basu 1992).  Depending on the details of the models,
the total CO and dust masses for the Galactic disk inferred at earlier
epochs are $\sim 2-5$ times larger than the current values.  If we
consider that some fraction of the Galactic halo mass ($\sim 2\times
10^{12} \msun$, Peebles 1995) is baryonic, such as low mass stars and
stellar remnants, the mass of CO and dust for the entire Galaxy may have
been more than an order of magnitude larger at the earlier epochs.

\section{Observational Studies}

In the previous section, we have calculated the evolution of the abundance
of CO, O$_2$, and dust in galaxies.  In this section we discuss the
feasibility of observing these species in galaxies at high
redshift.

\subsection{Observations of Dust at High Redshift}

Several researchers have already addressed the favorable prospects for
observations of dust at high redshift (Braine 1995; Eales \& Edmunds
1996), and many detections of dust in high redshift radio galaxies and
quasars have recently been reported (McMahon \etal 1994; Chini \&
Kr\"{u}gel 1994; Dunlop \etal 1994; Isaak \etal 1994; Ivison 1995; Omont
et al. 1996a).  For a universe with $q_o\sim0.5$ and $z>1$,
intrinsically similar sources observed at mm or sub--mm wavelengths have
observed flux densities which actually increase with increasing
redshift, due to the strong dependency of the dust emission on the
rest--frame frequency (McMahon \etal 1994).  Including the effects of
chemical evolution, we expect even brighter observed flux densities
because of the larger dust masses at earlier epochs
(\S\ref{sec-coo2dust}).  To quantify this statement, we calculate the
observed flux density $S_{\nu_{obs}}$ implied by the dust mass using the
expressions given by McMahon \etal (1994):
\begin{equation}
\label{eq:sdust}
M_{dust}=S_{\nu_{obs}}D_{L}^{2}[(1+z)\kappa_{d}(\nu_{r})B(\nu_r,T_{d})]^{-1},
\end{equation}
where $D_{L}$ is the luminosity distance which is given by
\begin{equation}
D_{L}=cH^{-1}_{o}q_{o}^{-2}(q_{o}z + (q_{o}-1)[(1+2q_{o}z)^{1/2}-1]),
\end{equation}
and $B(\nu_r,T_{d})$ is the Planck blackbody function for a rest--frame
frequency $\nu_{r}$ and dust temperature $T_{d}$.  For the absorption
coefficient of dust $\kappa_{d}$, we use a simple power law (Downes
\etal 1992):
\begin{equation}
  \kappa_{d}=\left\{ \begin{array}{ll} 0.4(\nu_r/250\GHz)^{2}
      \cm^{2}\g^{-1} & \nu_r < 1200 \GHz \\ 
      1.92 (\nu_r/250\GHz) \cm^{2}\g^{-1} & \nu_r > 1200
      \GHz. \end{array} \right.
\end{equation}
The dust absorption coefficient is uncertain, but it is expected to
increase with $\nu^{1}$ at the short infrared wavelengths and vary with
$\nu^{2}$ at longer wavelengths (Schwartz 1982; Hildebrand 1983).  

In Figure~\ref{ch_dust4} we show the evolution of the calculated thermal
dust flux density observed at 240~GHz (1.25 mm) as a function of
redshift.  The models in Figure~\ref{ch_dust4} are for galaxies with a
total final baryonic mass of $10^{12}\msun$.  By including evolution,
the computed flux densities are increased significantly with respect to
the nonevolutionary models.  For the closed box models, the maximum
intensity occurs before the maximum dust mass due to the increasing flux
density with increasing rest--frame frequency at high redshift (for
$q_{o}\sim 0.5$).  These effects are less important for the infall
models since the maximum amount of gaseous metals occurs at later
epochs.  The results presented in this paper suggest that observations
of dust emission from high--redshift young galaxies at mm and sub--mm
wavelengths are feasible.  Both the IRAM 30~m telescope bolometer system
(Wild 1995) and the Submillimeter Common User Bolometer Array [SCUBA] on
the James Clerk Maxwell Telescope (Matthews 1996) have sensitivities of
approximately 1--2 mJy for several hours of integration which is
sufficient to detect gas--rich galaxies with total masses of $M \ga
few\times 10^{11}\msun$ at high redshift.  The next generation of
mm--wavelength synthesis telescopes will achieve far greater
sensitivities of 0.1--0.2 mJy in only a minute of integration (Brown
1997) and will be very suitable for this research.

\subsection{Observations of Molecular Gas at High Redshift}

Despite many searches for CO emission in several different types of
suspected young galaxies, including QSOs, high redshift radio galaxies,
and damped \lya systems (Wiklind \& Combes 1994; van Ojik 1995;
Barvainis \& Antonucci 1996; Ivison \etal 1996; Frayer 1996; Evans \etal
1996; Yun \& Scoville 1997), the list of CO sources at high redshift is
still very limited.  Currently, the only three confirmed CO emission
detections at high redshift are for IRAS F10214+4724 (Brown \& Vanden
Bout 1991; Solomon \etal 1992), the Cloverleaf quasar (Barvainis \etal
1994a), and the quasar BR1202-0725 (Ohta et al. 1996; Omont et al.
1996b).  These systems have molecular masses of approximately
$10^{11}\msun$, assuming the Galactic CO--to--H$_2$ conversion factor
and assuming no gravitational amplification.  Both the Cloverleaf quasar
(Kayser \etal 1990) and IRAS F10214+4724 are lensed systems (Graham \&
Liu 1995; Broadhurst \& Lehar 1995).  It is still not known if lensing
is important for BR1202-0725.  Regardless of the precise interpretation
of these high--redshift systems, the observations indicate that these
systems have CO luminosities comparable to the most luminous infrared
galaxies seen at the current epoch (Downes, Solomon, \& Radford 1993;
Barvainis et al. 1995) and that copious amounts of metals were produced
very early in the history of the Universe ($t \la 1$~Gyr).

One of the goals of the models presented in this paper is to determine
the possibility of observing molecular oxygen during the early stages of
galaxy evolution.  Unlike CO, molecular oxygen is optically thin under
most ISM conditions (Black \& Smith 1985).  For the ground state
transition of $^{16}\oo [N(J)=1(1)\rightarrow 1(0)]$ which has a rest
frequency of 118.75~GHz, the line intensity in $\K \kps$ is given by
\begin{equation}
I(\oo) =6.8\times 10^{-16} N(\oo)/T,
\end{equation}
where $N(\oo)$ is the column density in $\cm^{-2}$ and $T$ is the
excitation temperature of the gas in Kelvin (L92).  The intensity
of the ground state O$_2$ transition relative to the \coa line is given
by
\begin{equation}
\label{eq:io2ico}
  \frac{I(\oo)}{I(\co)}=0.68\left(\frac{N(\oo)}{N(\co)}\right)
  \left(\frac{\alpha_{\co}}{\alpha_{\co,G}}\right)
\left(\frac{\xi_{\co}}{10^{-4}}\right) \left(\frac{30 \K}{T}\right),
\end{equation}
where $\alpha_{\co,G}$ is the Galactic CO--to--H$_2$ conversion factor
(e.g., Scoville \& Sanders 1987).  The global--averaged column density ratio
$N(\oo)/N(\co)$ corrected for relative filling factors is
\begin{equation}
\label{eq:fillf}
  \frac{N(\oo)}{N(\co)}=
\left(\frac{\oo}{\co}\right)_{dark}\times\left(\frac{M(\hh)^{\oo}}
  {M(\hh)^{\co}}\right)^{2/3},
\end{equation}
where $(\oo/\co)_{dark}$ is the abundance ratio within the dark clouds
and the ratio $M(\hh)^{\oo}/M(\hh)^{\co}$ is the volume--weighted mass
fractions given by equations (\ref{eq:fmh2co}\&\ref{eq:fmh2o2}).

An implicit assumption within equation (\ref{eq:io2ico}) is that both
the O$_2$ and CO lines are probing optically thick dense molecular
clouds.  For very powerful starbursts, the CO line emission may arise
predominantly from diffuse regions of low CO--optical depths near the
surfaces of molecular clouds (Aalto et al. 1995), while the O$_2$ lines
would probe only the dark cloud cores.  In such a scenario, the O$_2$/CO
line ratio would have little relationship to the actual global O$_2$/CO
abundance ratio.  However, it is still unclear if the $^{12}$CO emission
from starburst galaxies is actually dominated by regions of low optical
depth.  For the nearby starburst galaxy M82, Wild et al. (1992) find
that the $^{12}$CO emission arises from the optically thick dense
clouds, and in this case equation~(\ref{eq:io2ico}) would be applicable
after correcting for the relative CO and O$_2$ filling factors
(eq.[\ref{eq:fillf}]).  In addition, we expect the CO emission to arise
mainly from the cloud cores in metal--poor starbursts due to the lack of
shielding in the diffuse regions.

The Galactic CO--to--H$_2$ conversion factor is expected to vary with
metallicity in low metallicity systems, and at metallicities above solar
the conversion factor is expected to be constant since the CO emission
becomes optically thick and thermalized (Maloney 1990; Sakamoto 1996).
Using these assumptions, we adopt the following expression:
\begin{equation}
\label{eq:confac}
\left(\frac{\alpha_{\co}}{\alpha_{\co,G}}\right)  =\left\{
\begin{array}{ll} 
(Z/Z_{\odot})^{-0.7}& Z<Z_{\odot}\\
    1            & Z\geq Z_{\odot},
       \end{array} \right.
\end{equation}
where the power--law index of $-0.7$ has been observationally estimated
(Wilson 1995).  We emphasize that equation~(\ref{eq:confac}) has been
determined using a relatively small sample of low metallicity galaxies.
Significant deviations may exist from galaxy to galaxy or within any
particular galaxy.  For example, observations of clouds in the LMC and
SMC indicate a similar range of CO--to--H$_2$ conversion factors despite
their different metallicities (Rubio 1997).  These results suggest that
the conversion factor is not solely based on metallicity (e.g., see
appendix in Bryant \& Scoville 1996).

The mass of the H$_2$ molecules within the CO emission regions in
$\msun$ is related to the CO line intensity $I(\co)$ by
\begin{equation}
\label{eq:mh2co}
M(\hh)^{\co} h^{2}= 1.0 \times 10^{12} 
\left(\frac{\alpha_{\co}}{\alpha_{\co,G}}\right)
1.06\left(\frac{I_{\co}}{\mjy}\right)
\left(\frac{\Delta V}{\kps}\right)
\frac{q_{o}^{-4}Q^{2}}{\nu_{rest}^{2}(1+z)},
\end{equation}
where $\nu_{rest}$ is the rest frequency of the CO transition in GHz
[$\nu_{rest}=115.27$ for the \coa line] and $Q$ is a term associated
with the luminosity distance; $Q = q_{o}z +
(q_{o}-1)[(1+2q_{o}z)^{1/2}-1]$.  The factor of 1.06 arises from
assuming a Gaussian line shape with a FWHM line width of $\Delta V$.

In \S\ref{sec-coo2dust} we have calculated $M(\hh)^{\co}$.  Using
equation (\ref{eq:mh2co}), we calculate $I(\co)$, and then we determine
$I(\oo$) from equation (\ref{eq:io2ico}), assuming $T=30\K$.  For these
calculations, we adopt a FWHM line width of $300\kps$, which is
approximate CO line widths for the previous high--redshift CO
detections.  We show the results of the computed evolution of the \coa
and O$_2(1,1\rightarrow1,0)$ line intensities for a variety of models in
Figure~\ref{ch_imol4}.  These intensities may be scaled linearly with
the final total baryonic mass in units of $10^{12}\msun$, and the curves
scale inversely with line width.  The nonevolutionary solutions assume a
constant $M(\co)$ value which is the mass that is reached in the
numerical models at $t=10$~Gyr.

Currently, only the most massive or lensed galaxies can be detected in
CO emission.  With the next generation of radio telescopes, such as the
NRAO Green Bank Telescope (GBT), the Large Millimeter Telescope (LMT),
and the proposed mm--wavelength synthesis telescopes, redshifted CO
detections should be commonplace.  With several hours of integration
using these new instruments, we should be able to detect high--redshift
CO line strengths of $\sim 0.1-0.5 \mjy$ (Brown 1997).  This sensitivity
is sufficient to detect the \coa line in young massive galaxies
(Fig.~\ref{ch_imol4}).  Due to the $\nu^{2}_{rest}$ term in equation
(\ref{eq:mh2co}), observations of the higher transitions of CO
($J_{upper} >1$) can provide even more sensitive measurements of the
molecular gas mass, as long as the brightness temperature and the size
of the source do not decrease substantially compared to the \coa source
(Solomon \etal 1992).  Since starbursting galaxies at the current epoch
have similar brightness temperatures for the \coa and
CO(3$\rightarrow$2) lines (Devereux et al. 1994) and given that the
CO(6$\rightarrow$5) and CO(7$\rightarrow$6) lines have already been
detected at high--redshift, we expect equation (\ref{eq:mh2co}) to be
valid for at least the moderately high CO transitions, such as
CO(3$\rightarrow$2) and CO(4$\rightarrow$3), in high--redshift starburst
galaxies.  If this assumption is correct, the CO(3$\rightarrow$2) and
CO(4$\rightarrow$3) line intensities could be estimated by multiplying
the \coa curves in Figure~\ref{ch_imol4} by factors of 9 and 16,
respectively.  Even with the new instruments and observations of the
moderately high CO transitions, the Milky Way ($M_{T}\sim 10^{11}
\msun$) placed at high redshifts would still be very difficult to
detect, assuming no evolution.  However, the gas--rich predecessors of
galaxies similar to the Milky Way are expected to be observable.

The calculations presented in this paper suggest that both dust emission
and CO line emission will be readily observable at high redshift with
the next generation mm and sub--mm wavelength telescopes.  Observations
of molecular oxygen will be significantly more difficult.  We require
sensitivities of order 0.01~mJy for the O$_2$ lines at high redshift to
discriminate between the possible scenarios presented in this paper.
However, the predicted O$_2$ intensities greatly depend on many
uncertain parameters associated with photodissociation, molecular cloud
chemistry, and the densities and structure of molecular clouds.  For
simplicity, we have assumed uniform GMC densities.  If molecular clouds
are comprised of high--density clumps at low metallicities (Lequeux et
al. 1994), we may have underestimated the global O$_2$/CO mass ratios at
early epochs.  Given the theoretical uncertainties, observations of
O$_2$ in high--redshift CO sources could still be fruitful.  In addition
to high--redshift systems, metal--poor galaxies at the current epoch may
have enhanced O$_2$ abundances.  Ground--based observations of
metal--poor starbursts at $z\ga 0.03$ and observations of the LMC and
SMC with satellite telescopes, such as ODIN (Hjalmarson 1997), could
provide stringent constraints on the metallicity dependency of the
O$_2$/CO ratio.  Absorption line searches (\eg Combes \& Wiklind 1995)
along known low metallicity lines--of--sights could even be more
discriminating than the emission line studies.  Furthermore, new O$_2$
excitation calculations suggest the possibility of O$_2$ maser emission
for the N(J)=N(J)$\rightarrow$N(J$-1$) transitions within dense clouds
(Bergman 1995).  Without question, observations are required to
constrain the different theoretical hypotheses.

In this paper we have concentrated on the molecular lines.  For the high
rates of star formation expected in forming galaxies, the infrared
atomic lines which probe the photodissociation regions (Watson 1985) may
be significantly brighter than the molecular lines.  The redshifted
neutral atomic carbon transitions have been observed in both F10214+4724
(Brown \& Vanden Bout 1992) and the Cloverleaf (Barvainis \etal 1994b).
Searches for the redshifted C$^{+}$, N$^{+}$, and atomic oxygen lines,
which should all be stronger than the neutral carbon lines, are becoming
more feasible with instrumentation improvements.  With observations of
both the atomic lines and the molecular lines, we can constrain the
H$_2$/H ratios at high redshift and gain insight into the strength of
the photodissociation fields associated with young galaxies.  If the
majority of oxygen is not molecular, we would expect relatively strong
atomic oxygen emission lines from protogalaxies, given the expected
large O/C ratios at early epochs.  Since there are currently only a few
observations at high redshift, we cannot predict with any certainty
which atomic or molecular emission lines will be the strongest.  Future
observations will constrain the various possibilities.

\section{Conclusions}

By drawing upon the current knowledge in the fields of stellar
evolution, galactic chemical evolution, and chemical processes within
molecular clouds, we have constructed a general set of models to compute
the evolution of the abundance of CO, O$_2$ and dust as a function of
age and metallicity for massive galaxies.  Over a wide range of input
parameters, the models suggest the following general conclusions:
\begin{itemize}
\item The mass of gaseous metals ($ZM_g$) achieves its maximum at
  redshifts of $z\sim 1-3$ when the metallicity is approximately solar
  and when approximately $1/2$ of the galaxy mass is in gas.
\item The O/C ratio is expected to be $\ga 10$ at early epochs.
\item The observed increase in the C/O ratio with metallicity may
  suggest significant contributions to the yields from the winds of
  massive stars.
\item The O$_2$/CO ratio is expected to be of order unity in dark clouds
  ($A_{V}\ga 5$) during the early stages of galaxy evolution.
\item At the low metallicities, the global O$_2$/CO abundance ratios 
  corrected for photodissociation are are expected to be reduced by over an 
  order of magnitude relative to the dark cloud solutions.
\item The maximum CO and dust masses for young galaxies are expected to
  be at least 2--5 times larger than the values found for normal
  galaxies in the local universe.
\item The largest O$_2$ masses are expected for chemically young systems
  with metallicities of $0.1 < Z/Z_{\odot} < 1$ and large gas fractions
  ($\mu \ga 0.5$).
\item The models suggest that thermal dust emission and CO line emission
  should be observable for young massive galaxies at high redshift, and
  that even O$_2$ emission may be observable in chemically young
  galaxies.
\item The largest uncertainties in the models presented here are
  associated with the O$_2$ abundances.  The existence of detectable
  quantities of O$_2$ greatly depends on how metallicity,
  photodissociation, turbulent mixing, and grain processes affect the
  molecular gas--phase abundances.
\end{itemize}

In summary, all models predict more gaseous metals for young galaxies
since the mass of gaseous metals has been depleted in galaxies at the
current epoch due to star formation.  The models also indicate that the
abundances of the oxygen species, such as O$_2$, should be enhanced at
earlier epochs.  The models presented in this paper depend on many
uncertain parameters and approximations which require further
investigation.  Future observations of dust, CO, O$_2$, and the atomic
species will constrain the theoretical models and will greatly enhance
our knowledge of protogalaxies.

\acknowledgments

We thank S. Woosley and T. Weaver for their tables of supernova
ejecta.  We thank A. Maeder for communications concerning the
yields in M92.  We are also very grateful to J. Black for several helpful
comments and suggestions.  DTF was supported by the NRAO predoctoral
program and by a postdoctoral fellowship funded by the University of
Toronto.

\clearpage
% Figure Captions for ApJ submission

\figcaption[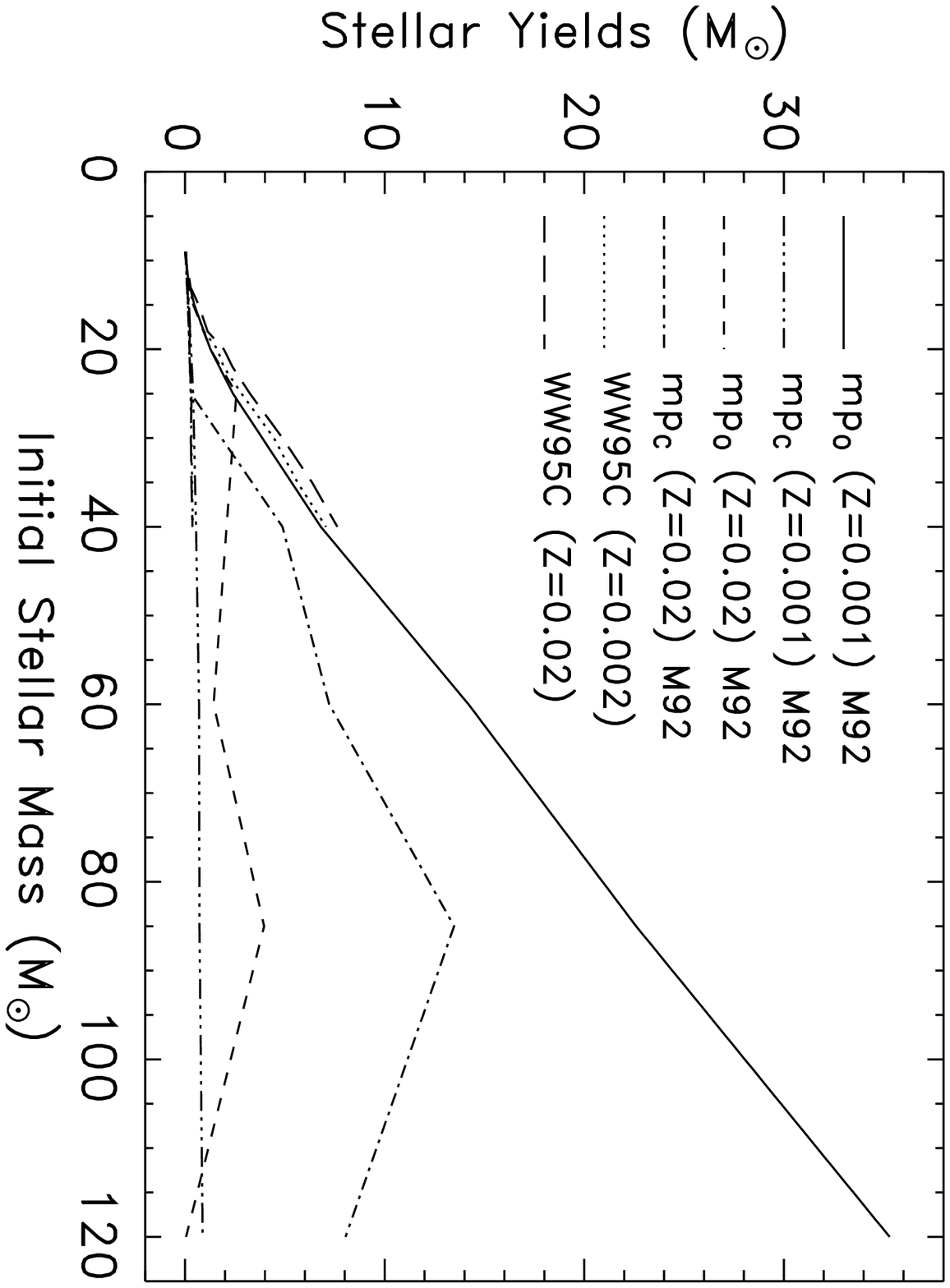]{The stellar yields of oxygen $(mp_O)$ and carbon $(mp_C)$
  calculated by Woosley \& Weaver (1995) and Maeder (1992) for solar and
  subsolar metallicities.  The yields of M92 and WW95C are consistent at
  low metallicities, but for solar metallicities their yields diverge
  for high mass stars due to the inclusion of mass loss via stellar
  winds in the M92 models. \label{yields}}

\figcaption[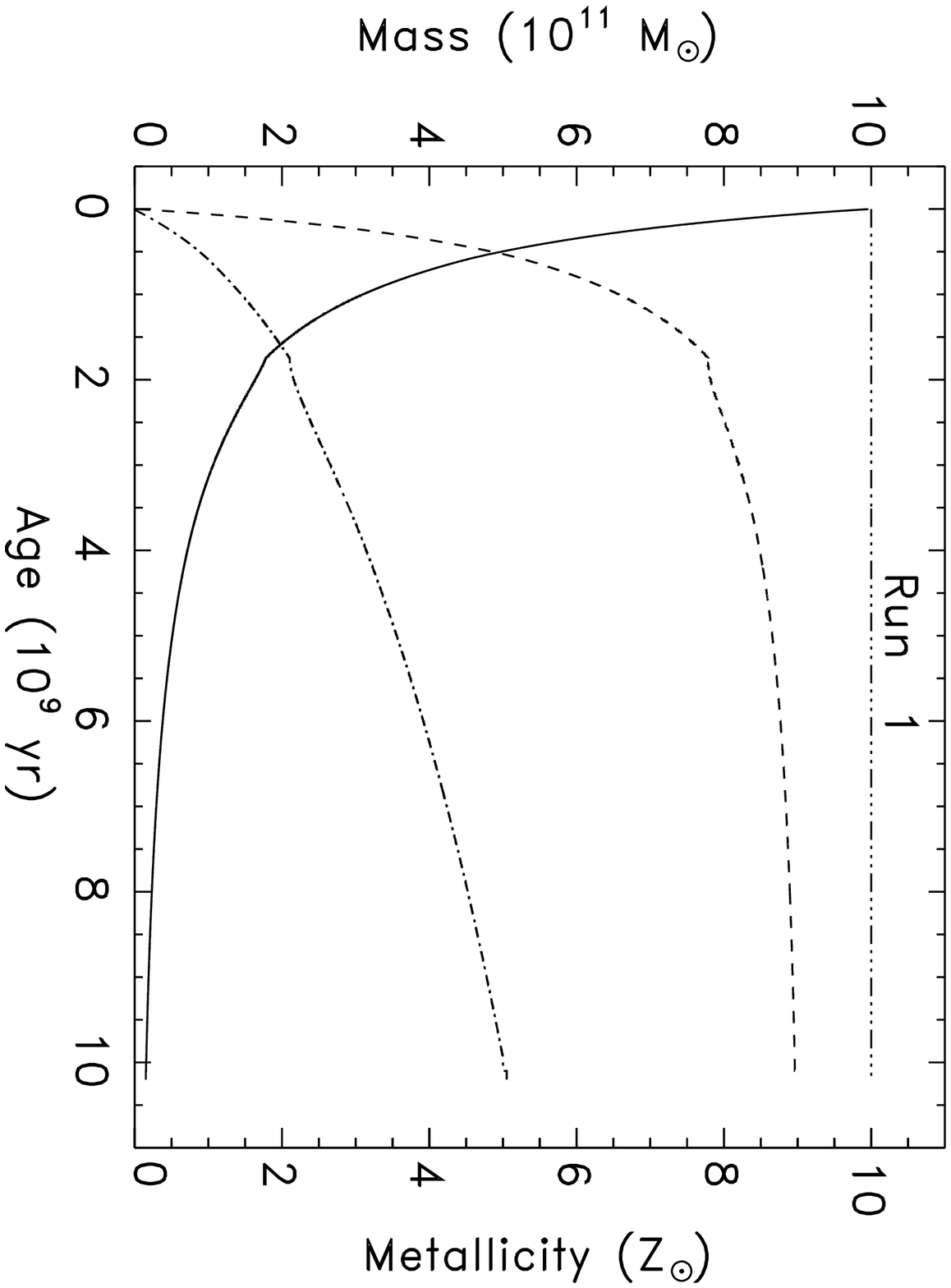]{The evolution of gas mass ($M_g$; solid line), stellar mass
  ($M_S$; dashed line), total mass ($M_{T}$; dash--dot--dot--dot line),
  and metallicity (dash--dot line) as a function of age for a typical
  closed box model (Run~1). The mass in stellar remnants is $M_r=M_T -
  M_g -M_S$. \label{evolt1}}

\figcaption[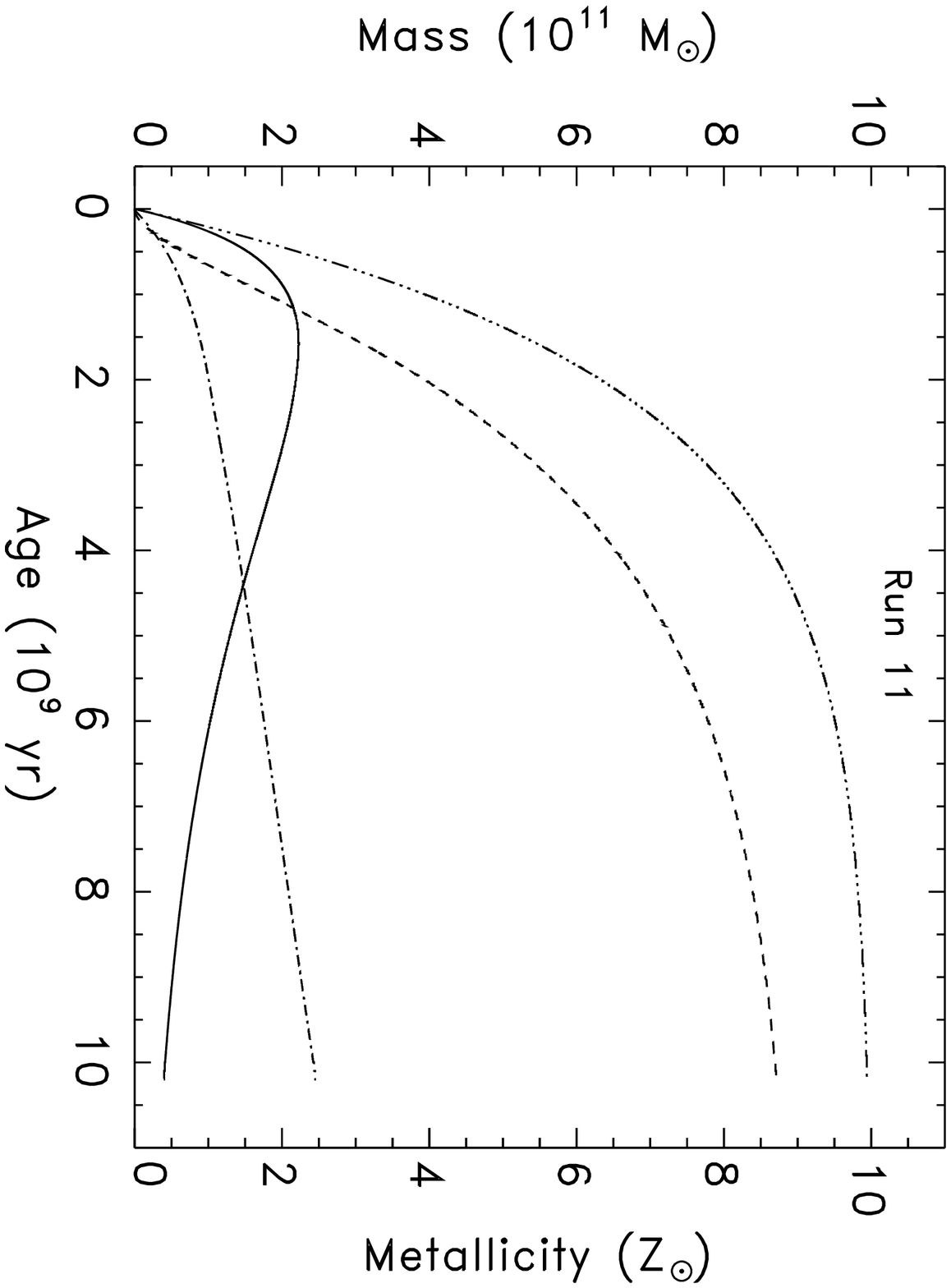]{The evolution of gas mass (solid line), stellar mass
  (dashed line), total mass (dash--dot--dot--dot line), and metallicity
  (dash--dot line) as a function of age for a typical infall model
  (Run~11). \label{evolt11}}

\figcaption[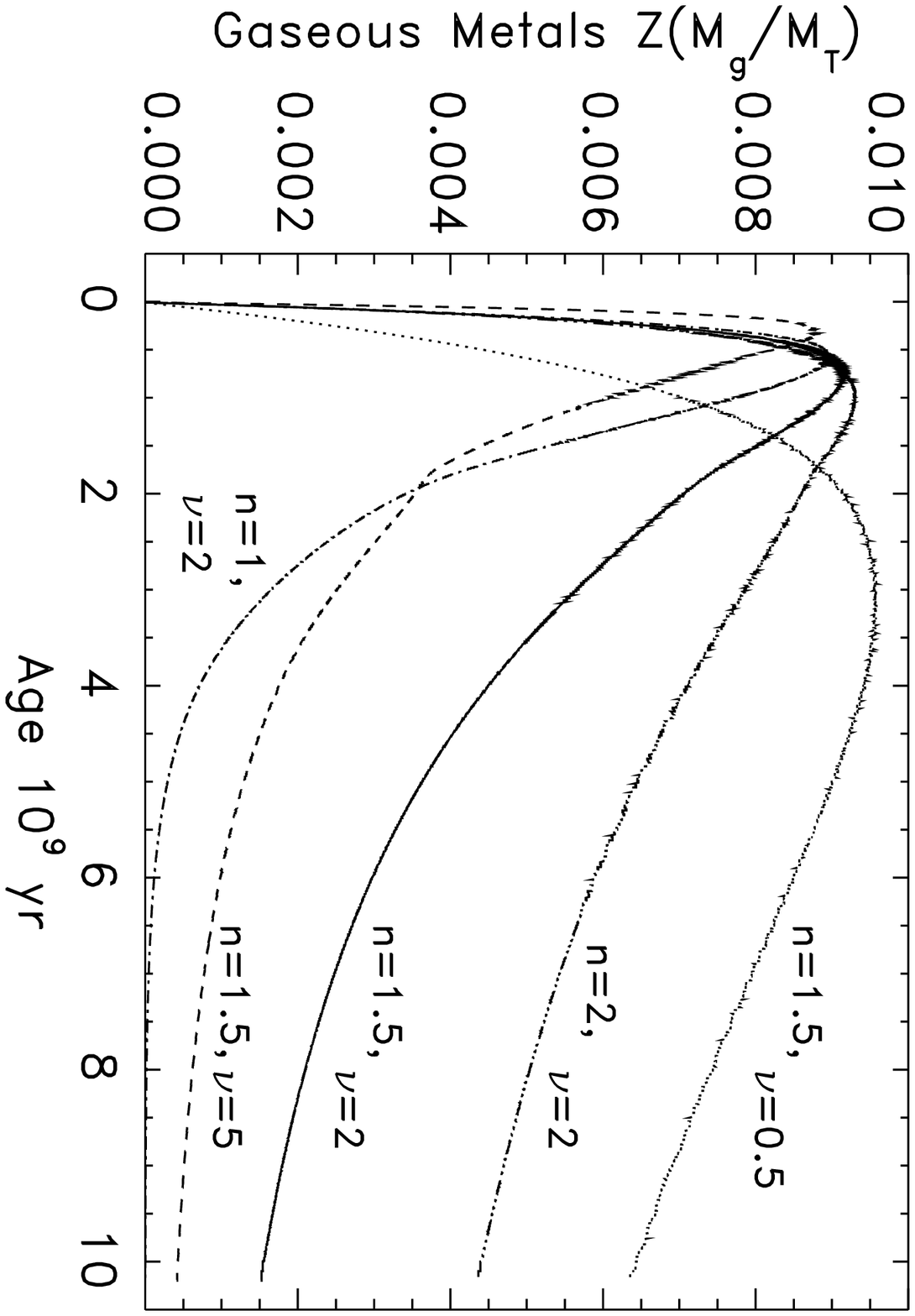]{The evolution of gaseous metals $(Z M_g)$ for
  different star formation rate parameters.  These curves are derived
  using the yields of M92 ($m_u=120\msun$), using $x=1.35$, and with no
  infall (closed box models). \label{zmgm92}}

\figcaption[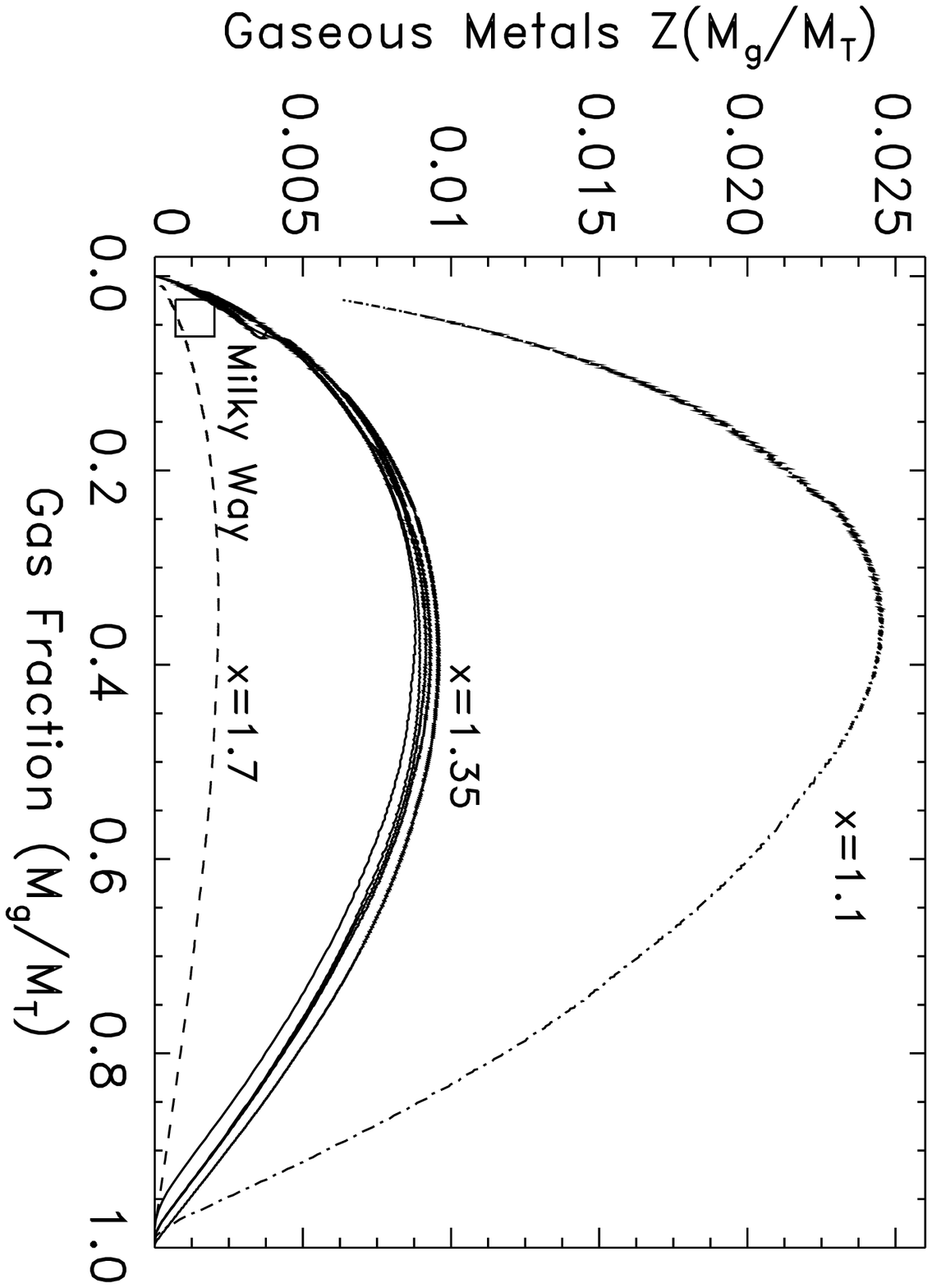]{The evolution of gaseous metals as
  a function of the gas fraction for different IMFs.  These curves are
  solutions of closed box models using the yields of M92.  The central
  group of curves represents all solutions plotted in
  Figure~\protect\ref{zmgm92}.  \label{muzm92} }

\figcaption[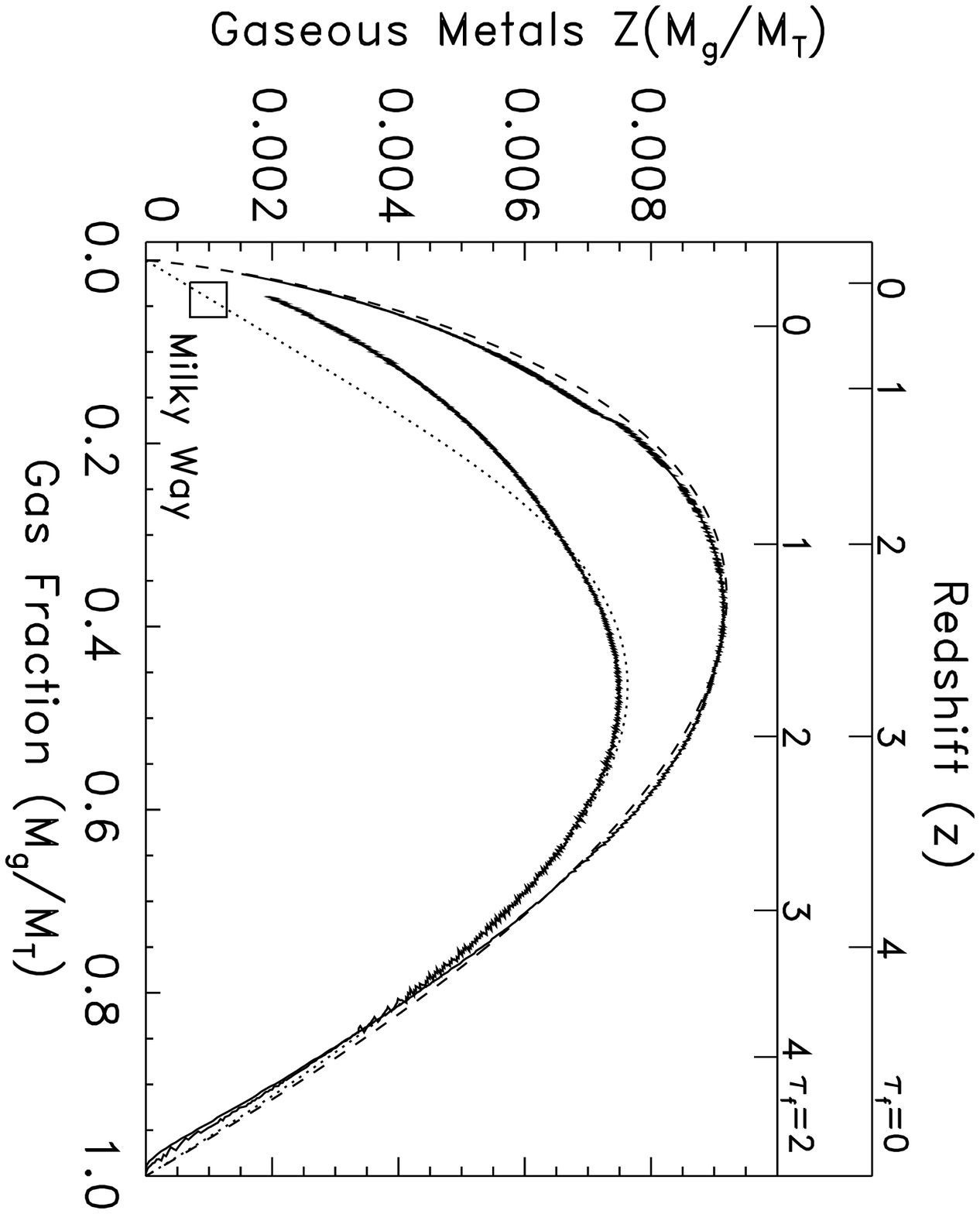]{A comparison of the numerical solutions with the
  analytical solutions. The upper and lower redshift scales are for
  numerical solutions of Run~1 and Run~11, which are represented by the
  upper and lower solid lines, respectively.  The analytical closed box
  solution is the dashed line (eq.~\protect[\ref{eq:cbox}]) which is
  similar to the closed box numerical solutions.  The analytical simple
  infall solution is the dotted line (eq.~\protect[\ref{eq:sinfall}])
  and is similar to the numerical infall solutions.  As $f(t)\rightarrow
  0$ the numerical infall solutions approach the closed box solutions.
  All models predict that the peak of gaseous metals as a fraction of
  the total mass occurs when approximately $1/2$ of the total mass is in
  stars and $1/2$ is in gas.  The redshift scales assume $h=0.8$,
  $q_o=0.5$, and $z_f=5$.
  \label{paul}}

\figcaption[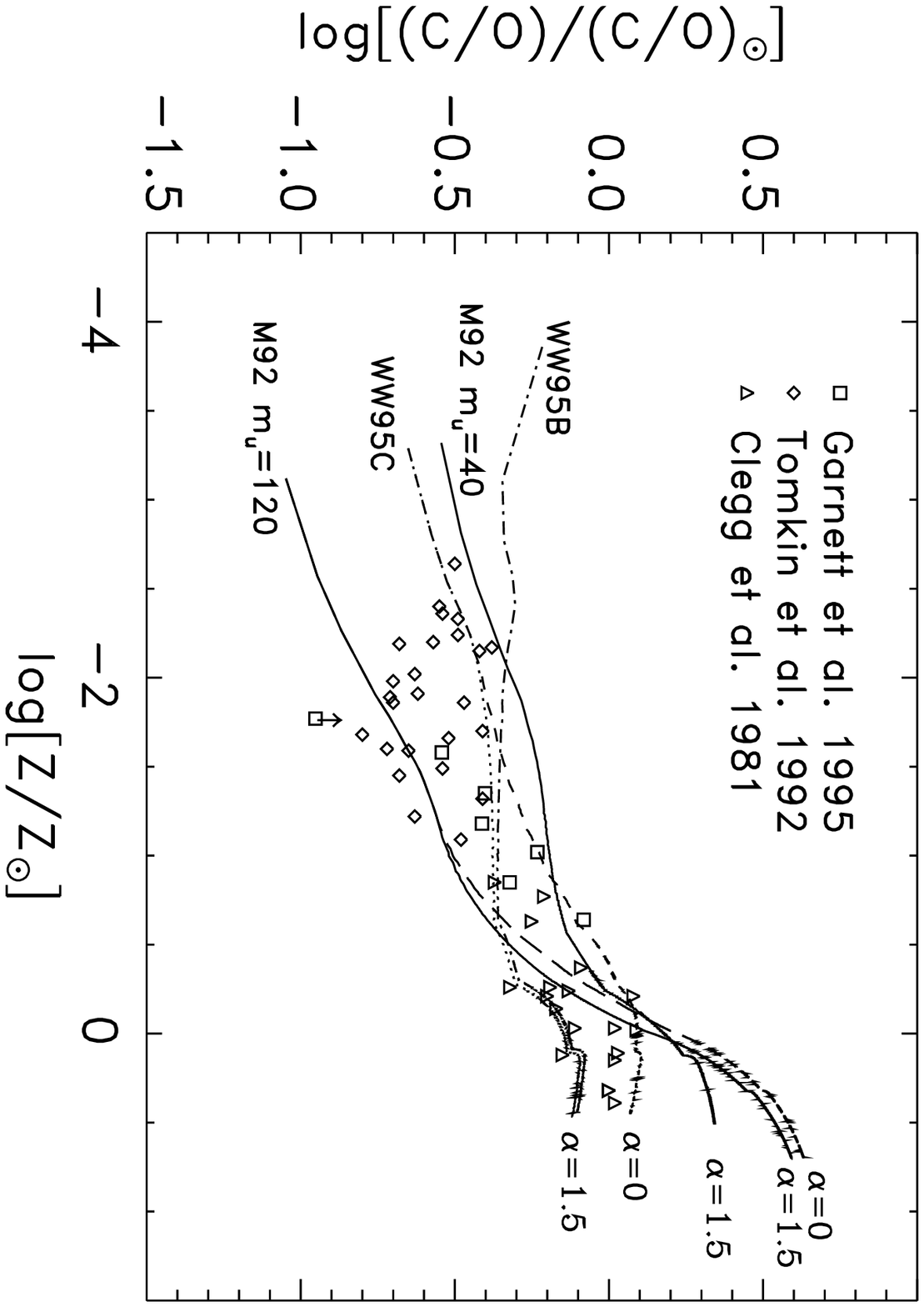]{The evolution of the C/O ratio as a function of metallicity
  for different stellar yields, where the boxes represent abundance
  measurements of {\sc Hii} galaxies, the diamonds are for halo stars
  and the triangles are for disk stars.  All curves use $x=1.35$,
  $\nu=2$, $n=1.5$, and $\tau_f=0$.  The stellar yields of M92 better
  fit the observed increase of the C/O ratio with metallicity and 
  indicate the significance of stellar winds from massive stars which
  were neglected in the WW95 models. \label{coz}}

\figcaption[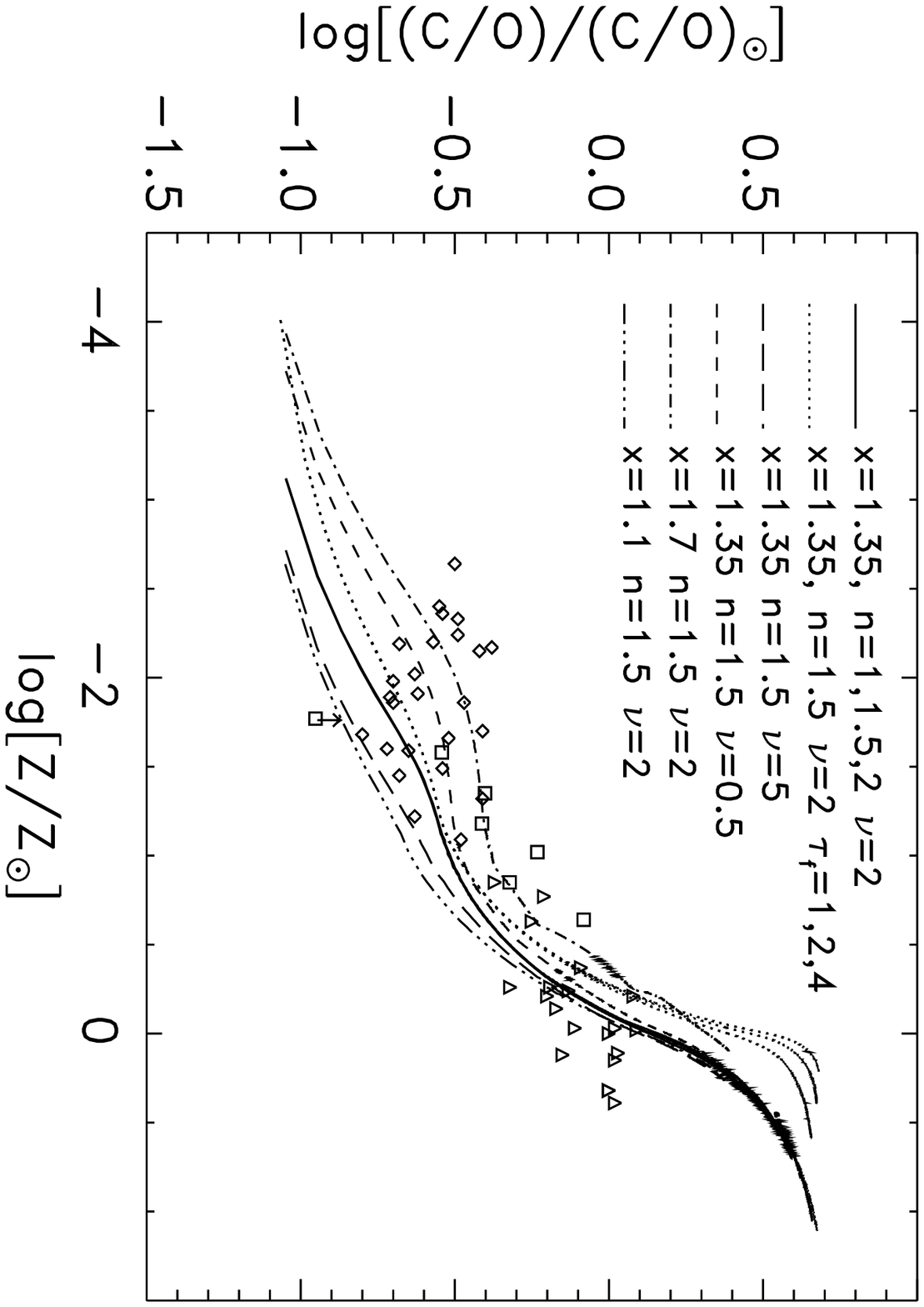]{The variation of the C/O ratio as a function of
  metallicity for different star formation rates, IMFs, and infall
  parameters using the yields of M92.  The boxes, diamonds, and
  triangles represent the same observations shown in
  Figure~\protect\ref{coz}. \label{cozm}}

\figcaption[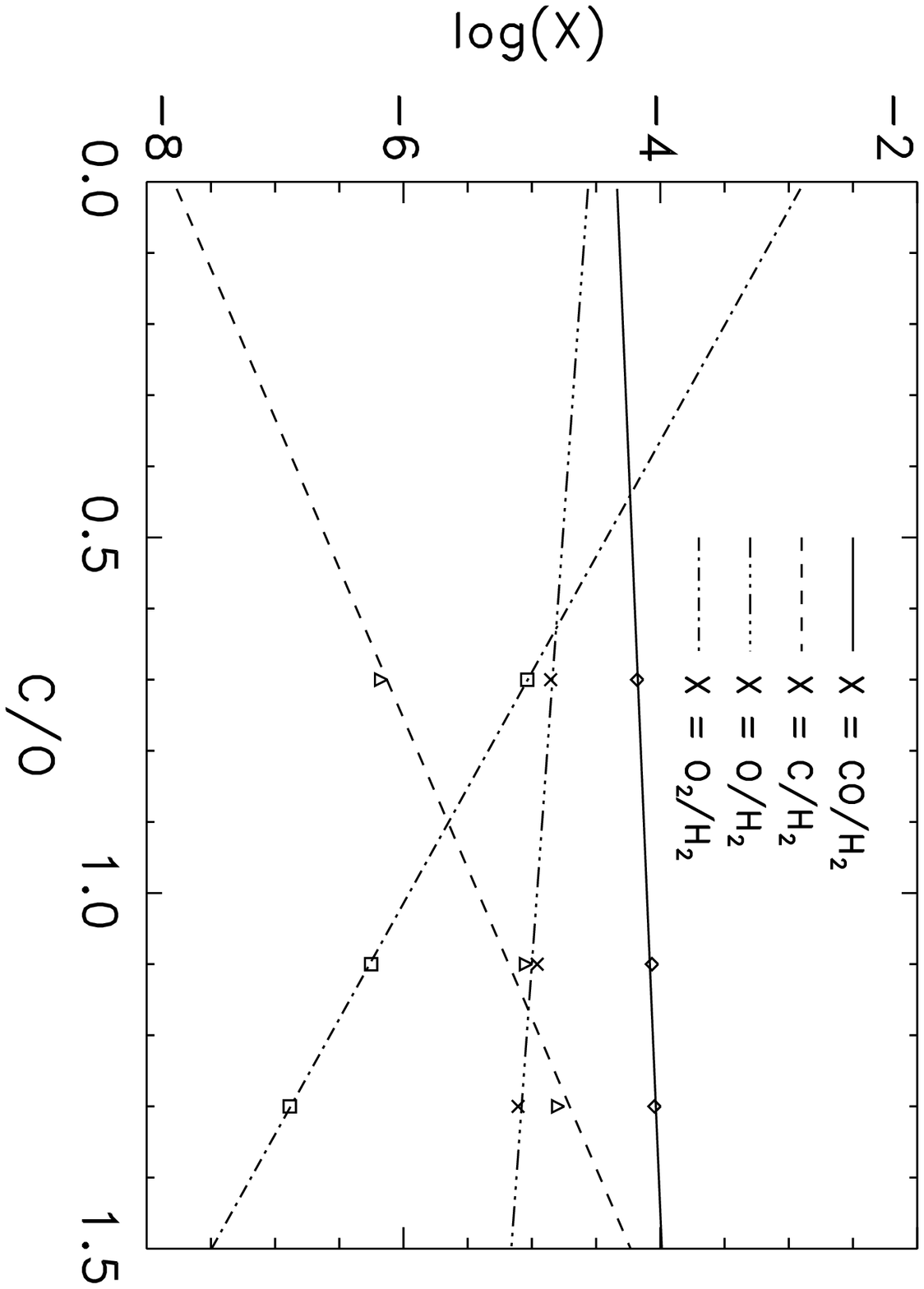]{Fits to the molecular cloud abundances.  The data
  points are gas--phase steady--state solutions for the dark cloud
  models of LG89.  These data demonstrate that the abundances are
  extremely sensitive to the gas phase C/O ratio. \label{lg89fit}}

\figcaption[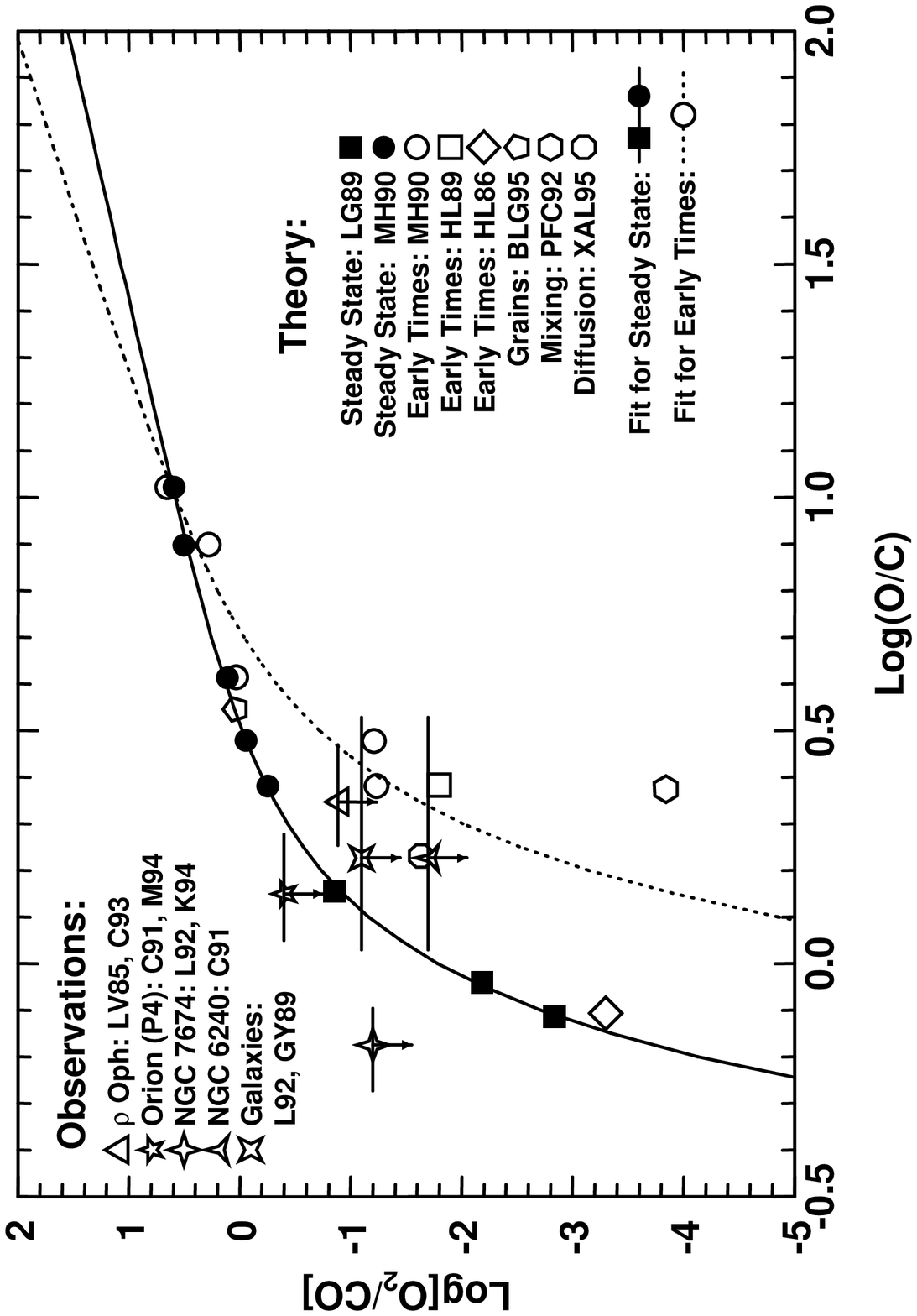]{The O$_2$/CO ratio as a function of the O/C ratio.
  The $2\sigma$ observational upper limits of the O$_2$/CO ratio are
  shown along with their estimated uncertainties of the O/C ratio.  The
  theoretical data are from many different time dependent molecular
  chemistry models.  For O/C$\ga 10$ the solutions at early times
  converge with the steady--state solutions.  Molecular clouds may be
  described by the early--time solutions which could result from
  turbulent mixing of the gas and may explain the apparent lack of O$_2$
  at solar abundances [$(\io/\ac)_{\odot} \simeq 2$].
  \label{lgmh}}

\figcaption[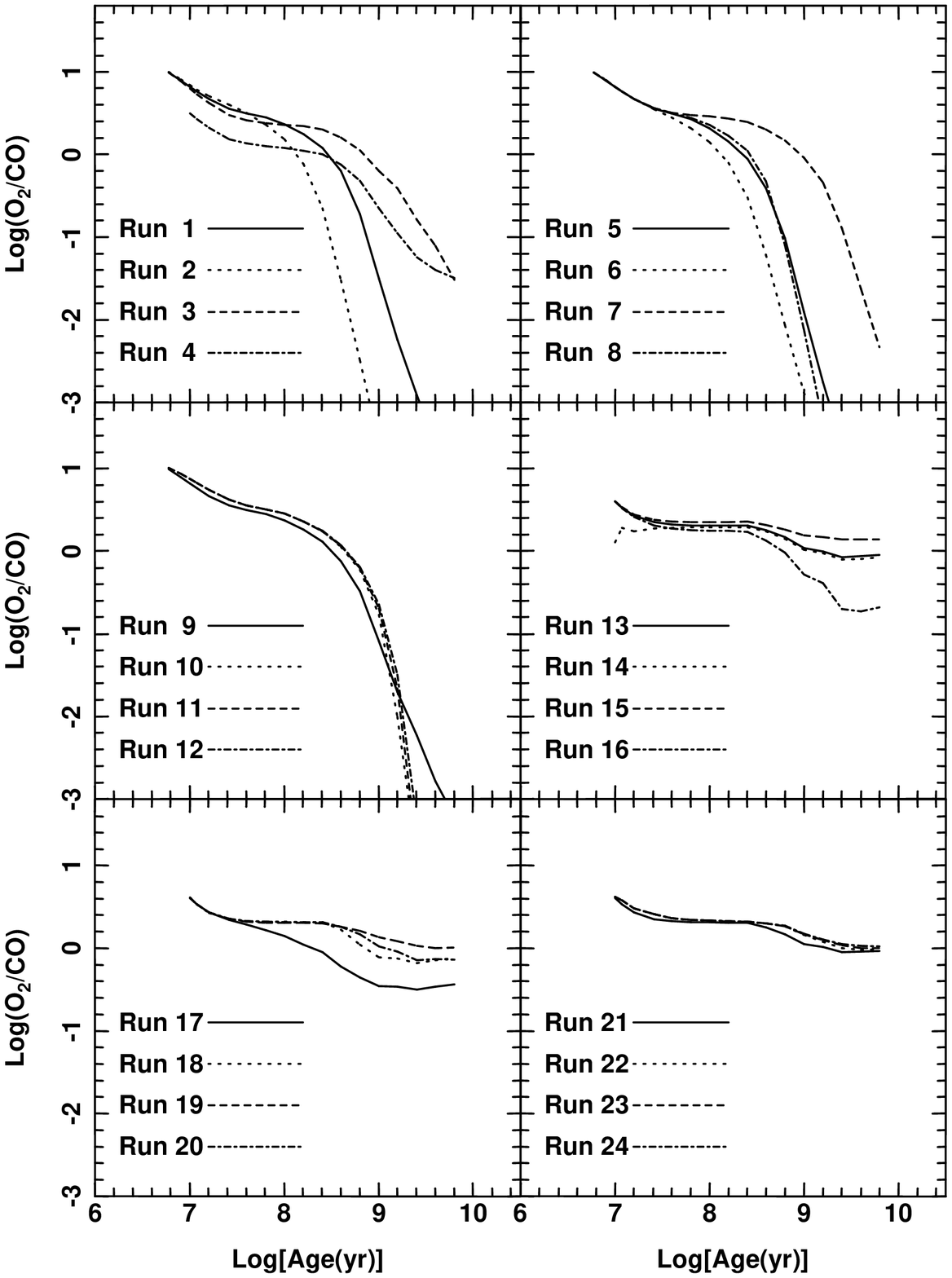]{The evolution of the gas--phase steady--state O$_2$/CO
  abundance ratio within dark molecular clouds ($A_{V} \ga 5)$ as a
  function of age for all models. \label{o2cotall}}

\figcaption[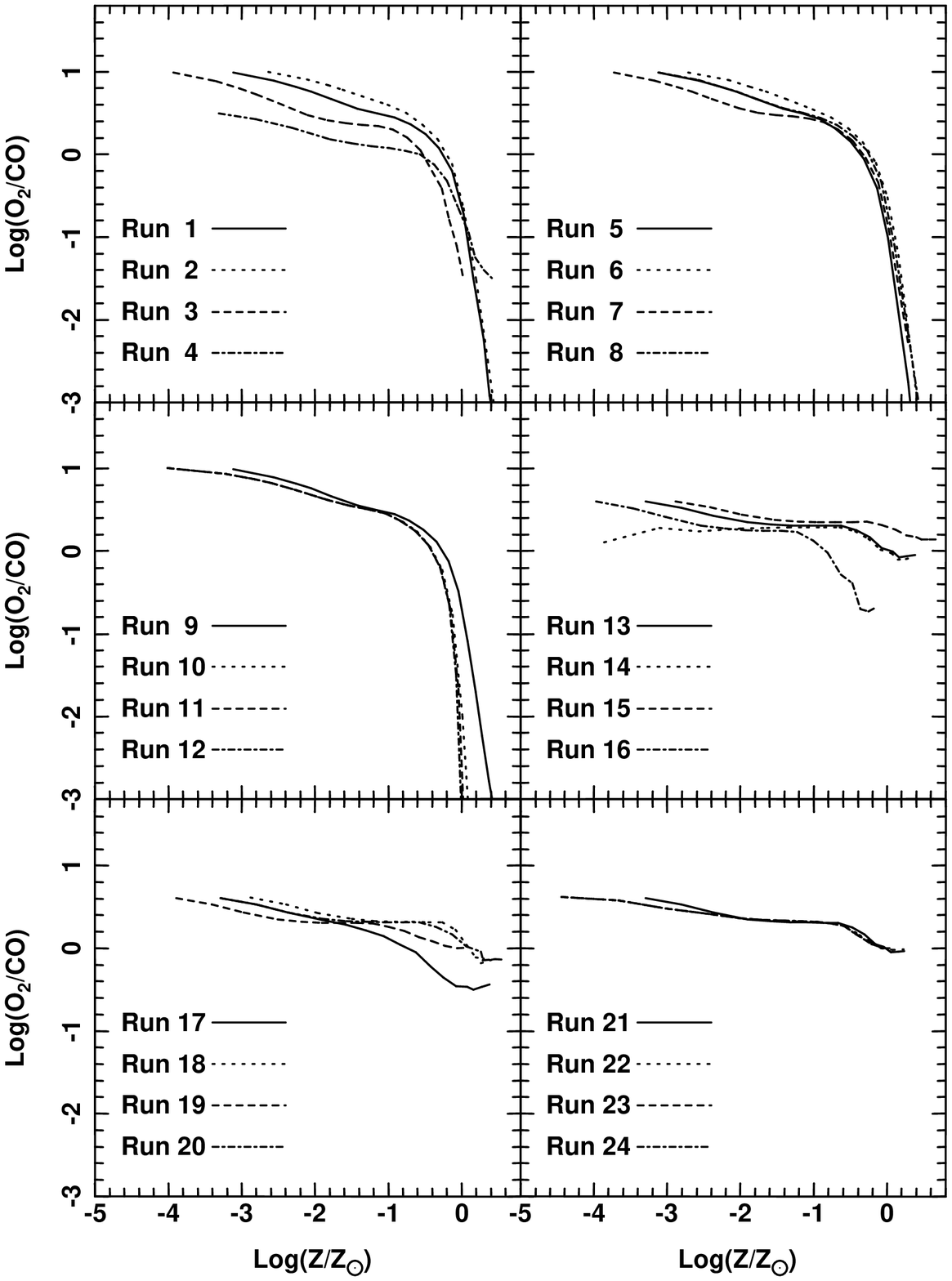]{The evolution of the gas--phase steady--state O$_2$/CO
  abundance ratio within dark molecular clouds as a function of
  metallicity for all models. \label{o2cozall}}

\figcaption[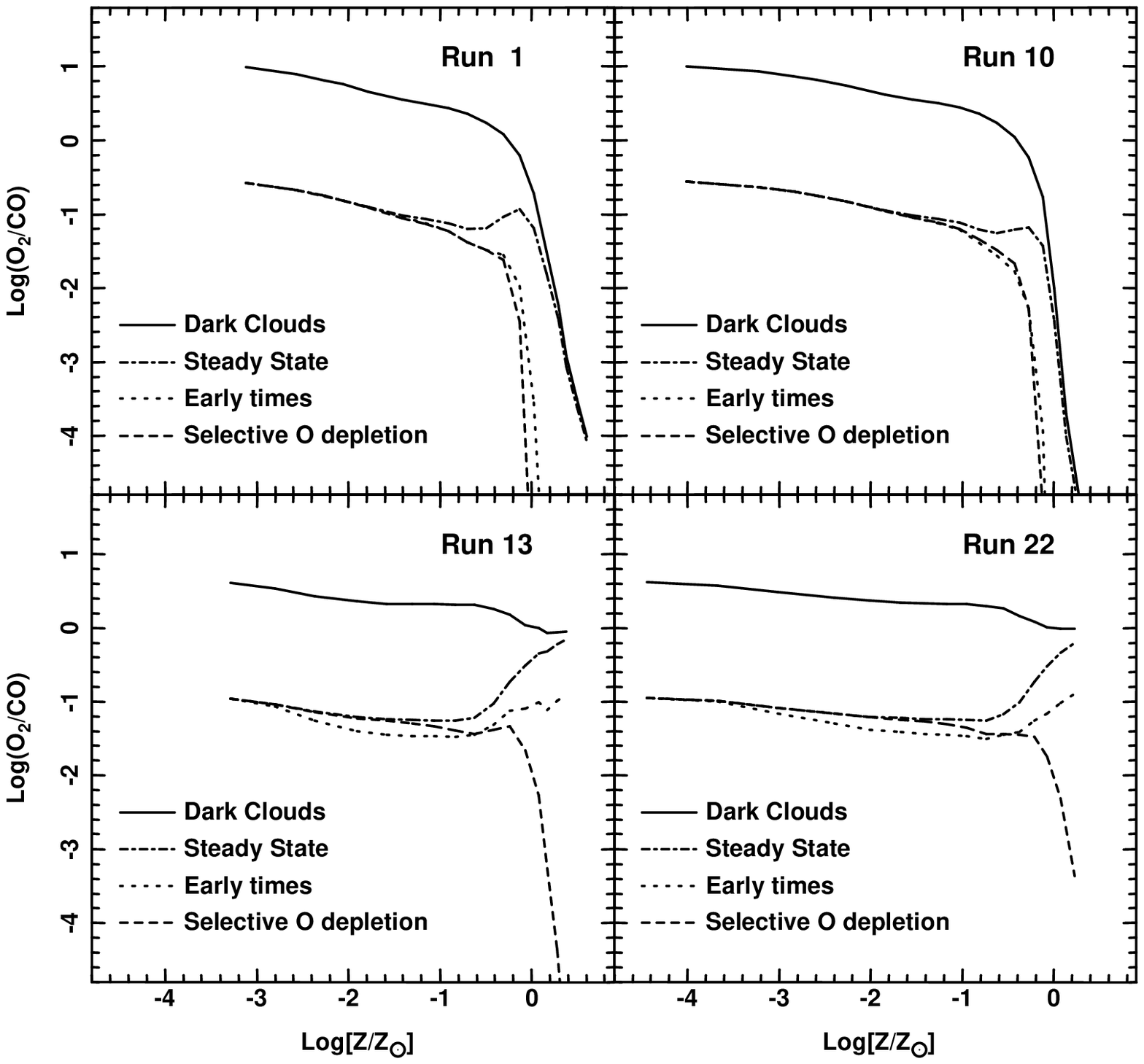]{The evolution of the global volume--weighted
  O$_2$/CO abundance ratio corrected for photodissociation effects as a
  function of metallicity and the SFR (see text).  The solid line
  represents the uncorrected steady--state solutions which are only
  applicable for dark molecular clouds.  Run~1 and Run~13 are closed box
  models using the stellar yields of M92 and WW95C, respectively.
  Run~10 and Run~22 are infall models using the stellar yields of M92
  and WW95C, respectively.
  \label{ch_mol4}}

\figcaption[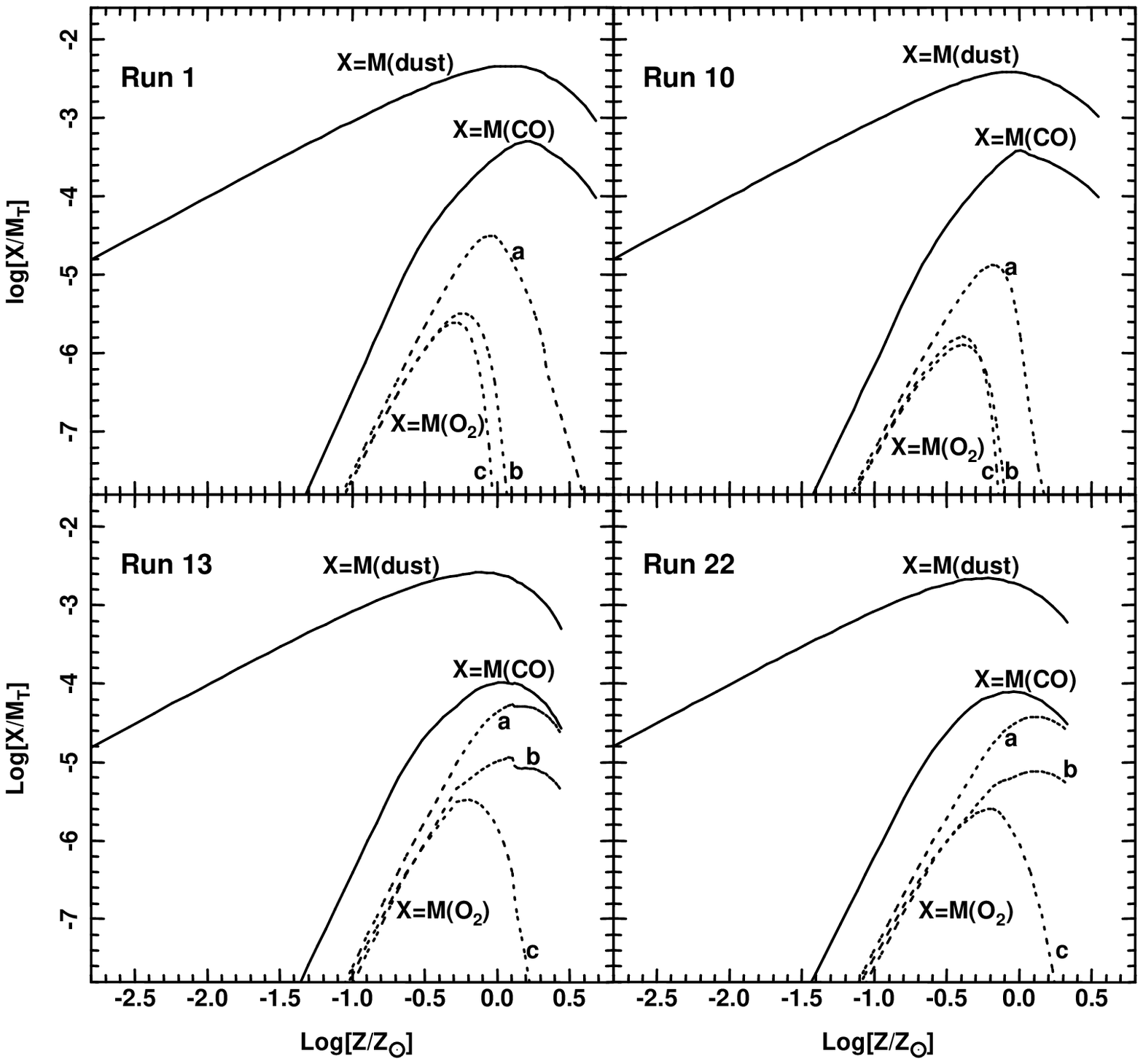]{The evolution of the dust, CO, and O$_2$
  (short--dashed lines) masses as a function of metallicity. The models
  are the same as those shown in Figure~\protect\ref{ch_mol4}.  The
  three scenarios for the evolution of O$_2$ (volume--weighted
  abundances corrected for photodissociation effects) are labeled as (a)
  steady state, (b) early times, and (c) selective O depletion (see
  text). \label{ch_z4}}

\figcaption[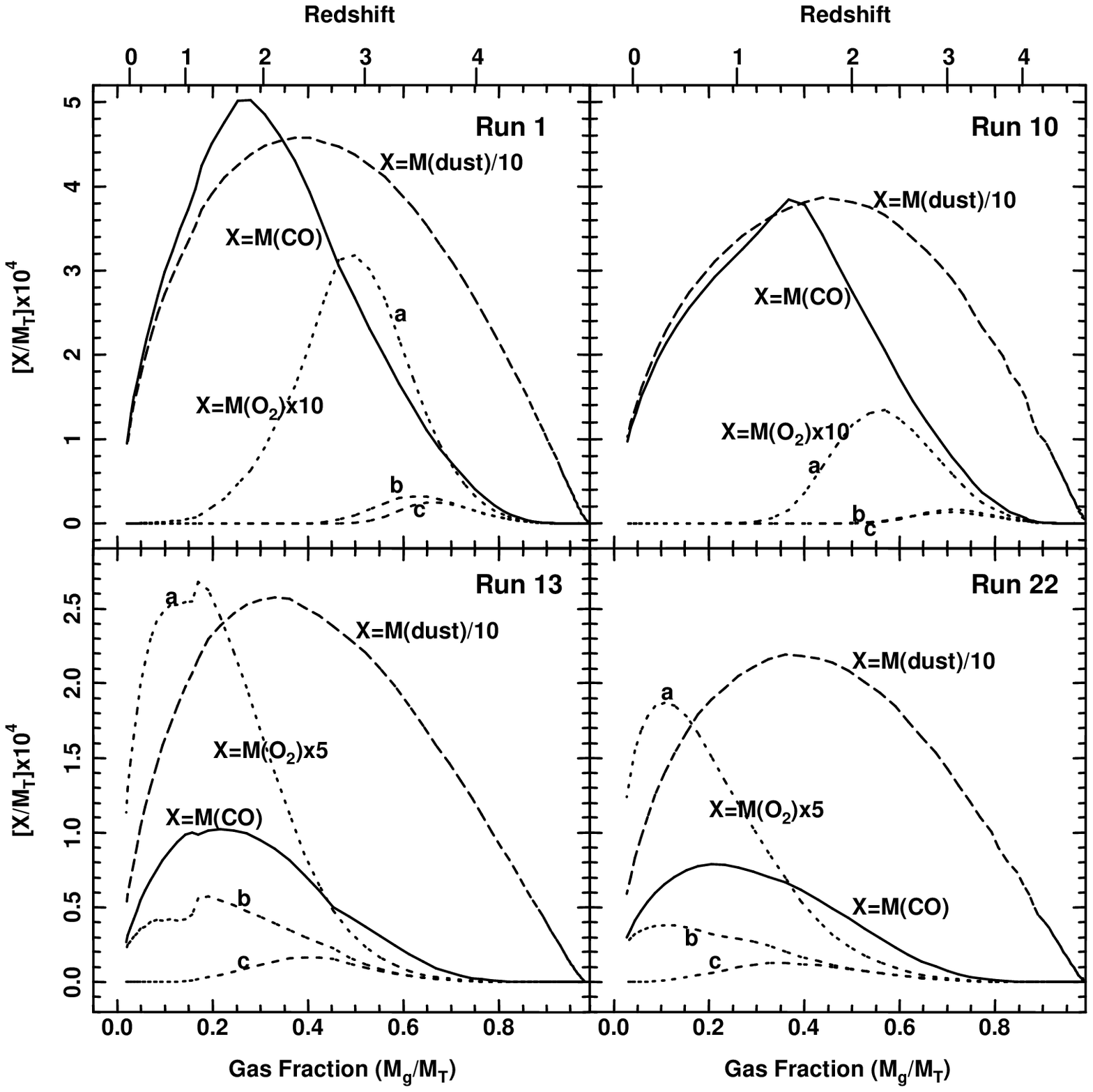]{The evolution of the dust mass (long--dashed line),
  CO mass (solid line), and O$_2$ mass (short--dashed lines) as a
  function of the gas fraction.  The models are the same as those shown
  in Figure~\protect\ref{ch_z4}.  The redshift scales assume $h=0.8$,
  $q_o=0.5$, and $z_f=5$.  \label{ch_mu4}}

\figcaption[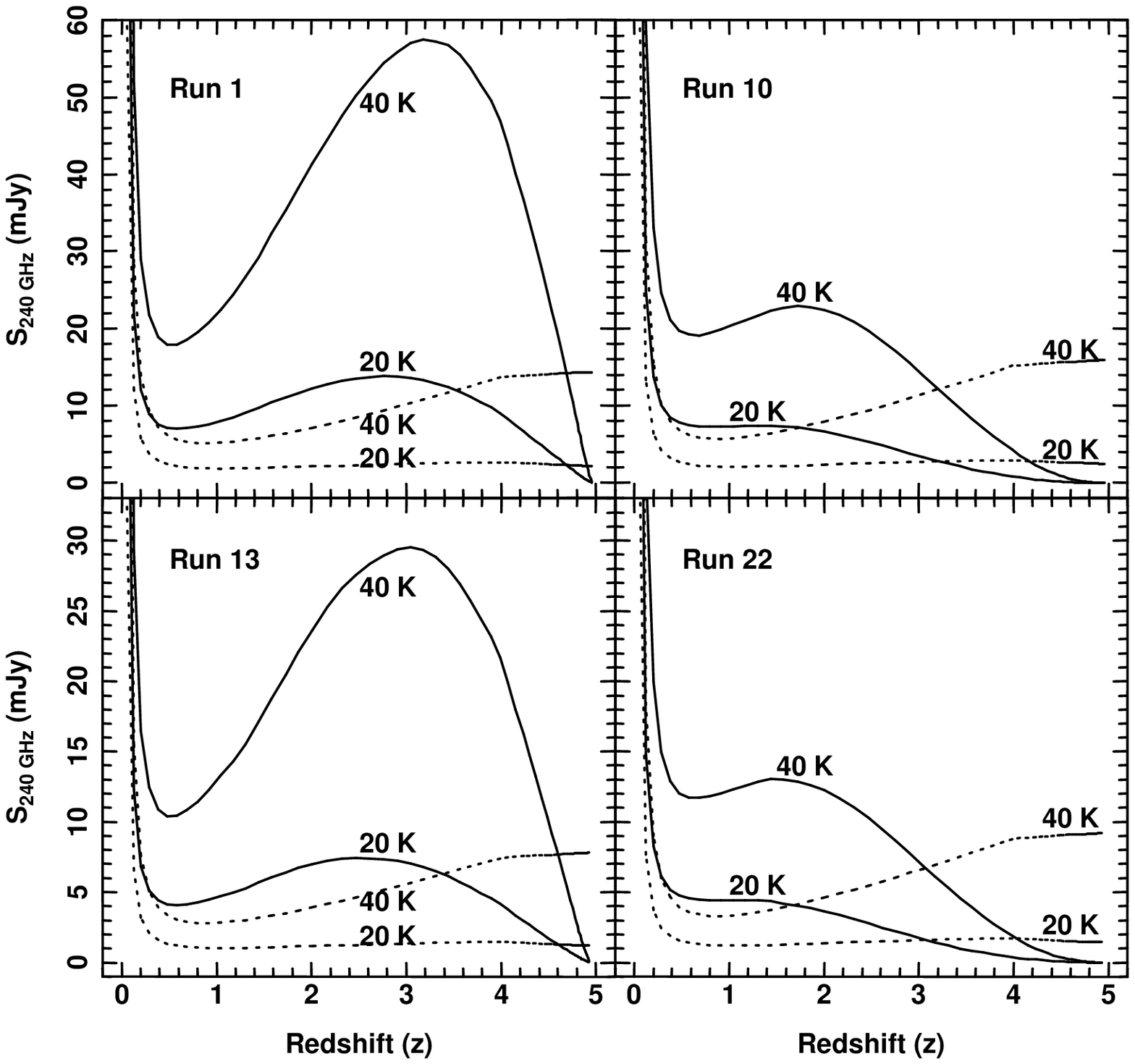]{The evolution of the thermal dust continuum emission
  observed at 240~GHz (1.25 mm) as a function of redshift for galaxies
  with a final total baryonic mass of $10^{12}\msun$ and dust
  temperatures of 40~K and 20~K (eq. [\ref{eq:sdust}]).  The solid lines
  are calculations for the same evolutionary models and cosmological
  parameters used in Figure~\protect\ref{ch_mu4}.  The dashed lines
  represent nonevolutionary scenarios with a dust mass achieved at
  $t=10$~Gyr in their corresponding numerical model.
  \label{ch_dust4}}

\figcaption[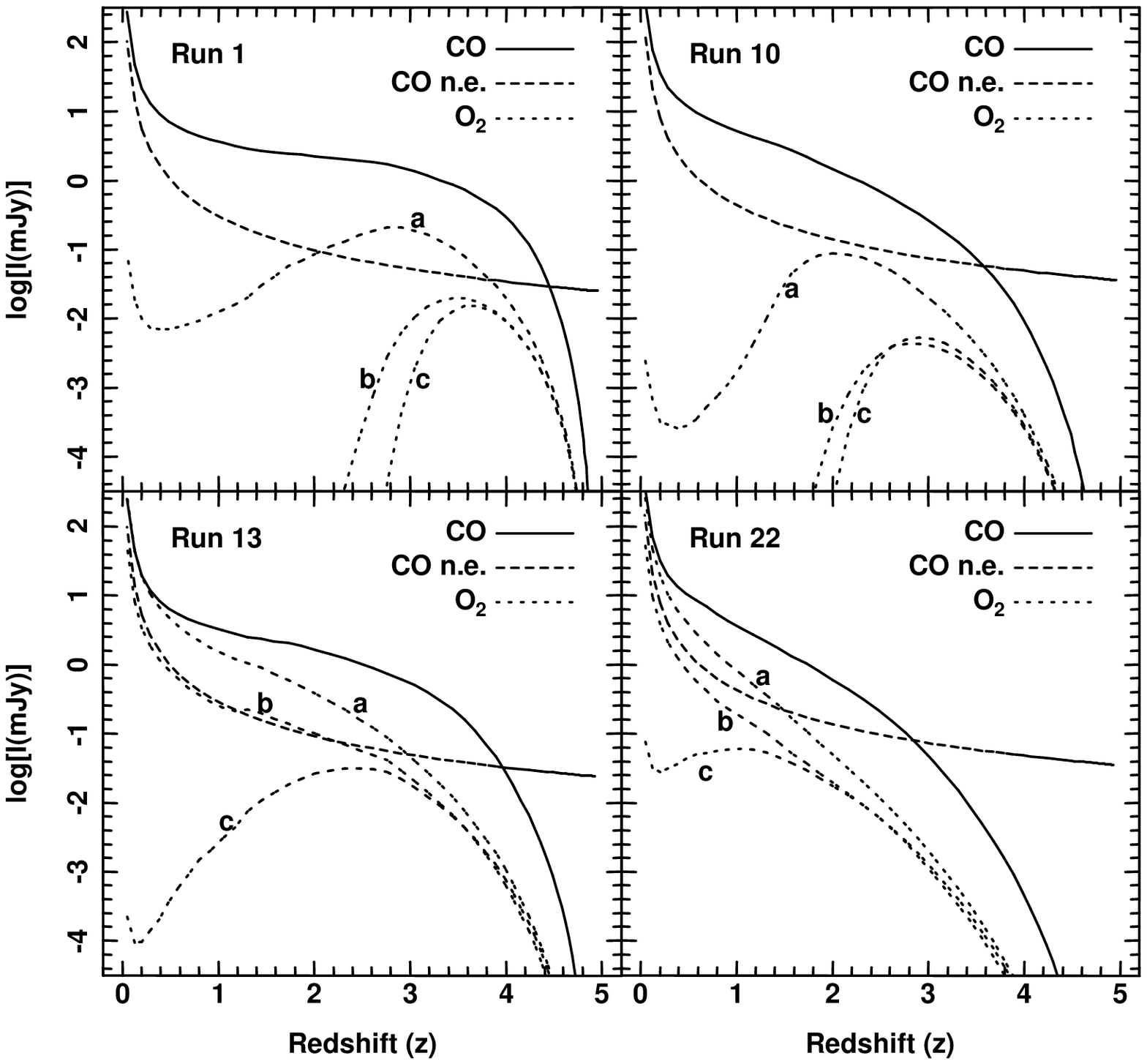]{The evolution of the strengths of the \coa and
  O$_2(1,1\rightarrow 1,0)$ line intensities as a function of redshift
  for galaxies with a final total baryonic mass of $10^{12}\msun$ and
  with a FWHM line width of $300\kps$.  The calculations are for the
  same evolutionary models and cosmological parameters used in
  Figure~\protect\ref{ch_mu4}.  The nonevolutionary scenarios (CO n.e.;
  long--dashed lines) are for the total mass of the CO emission regions
  achieved at $t=10 \Gyr$ in their corresponding numerical model.
  \label{ch_imol4}}

%\end{document}

%ApJS staff please ignore the lines below which allows us to include
%the ps-files containing figures and tables for preprints.

\input tabs.tex
\input figs.tex

\end{document}

%% file: tabs.tex
%Include tables as ps-files to maintain proper spacings
\newpage
\plotone{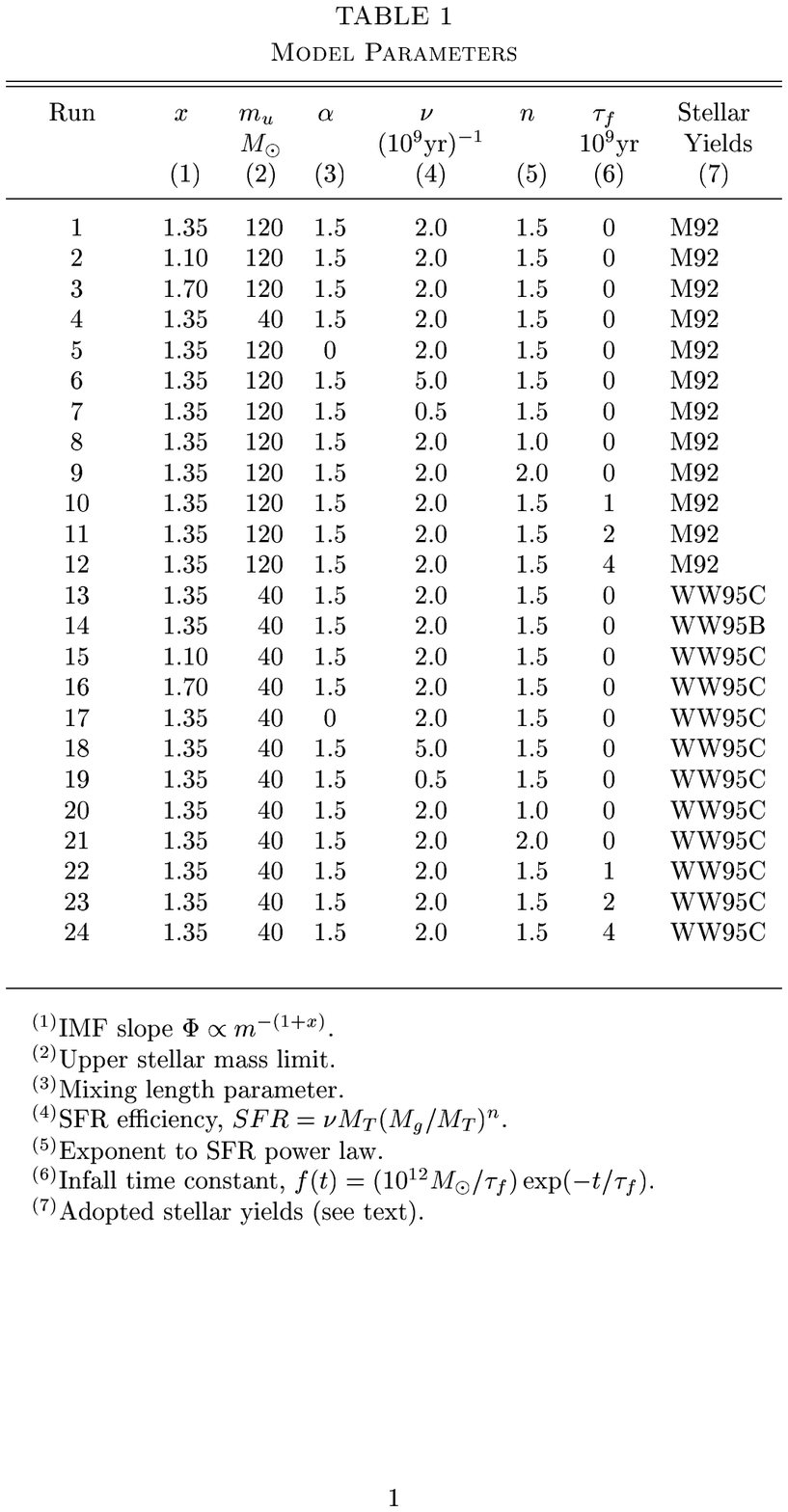}
\newpage
\plotone{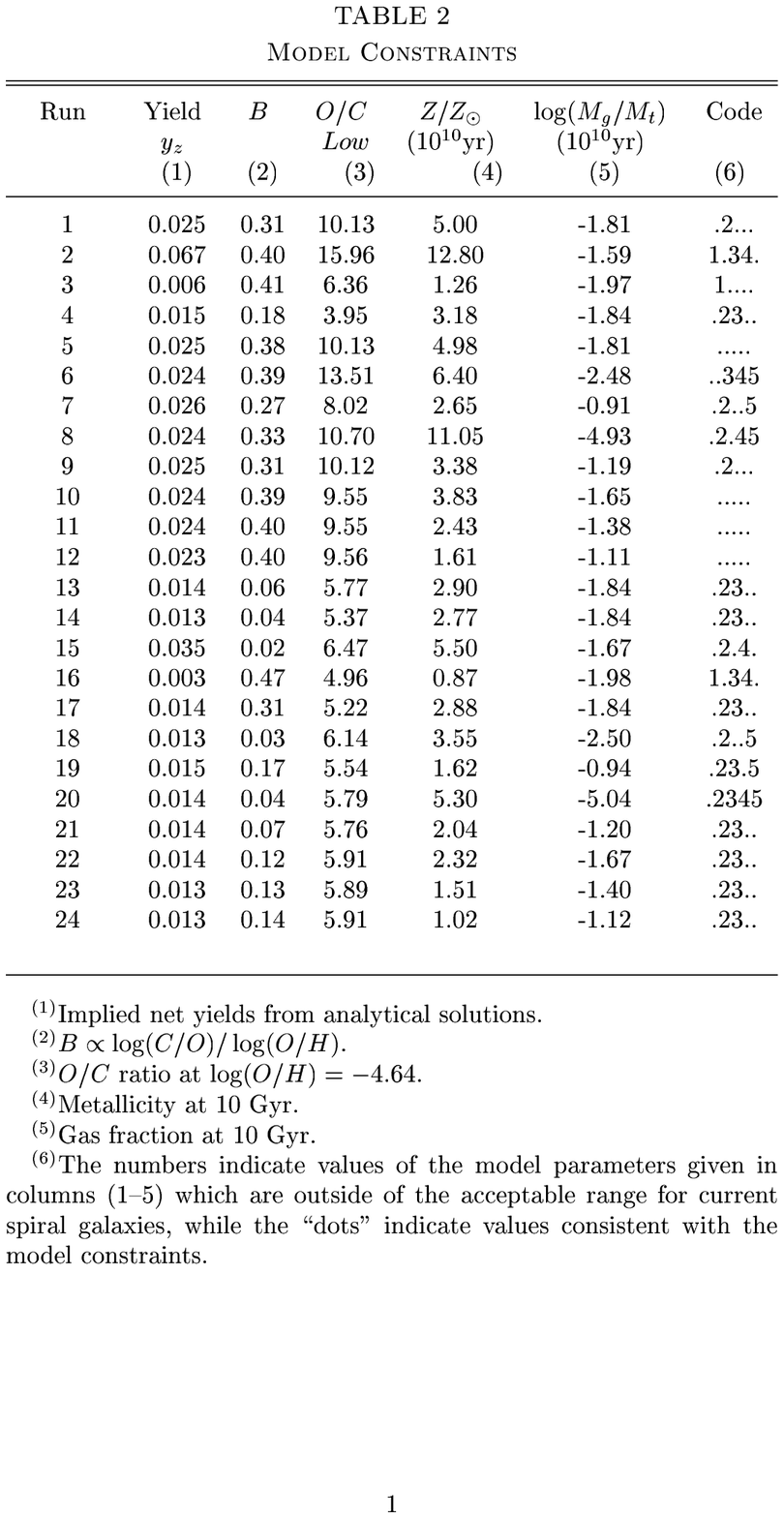}
\newpage
\plotone{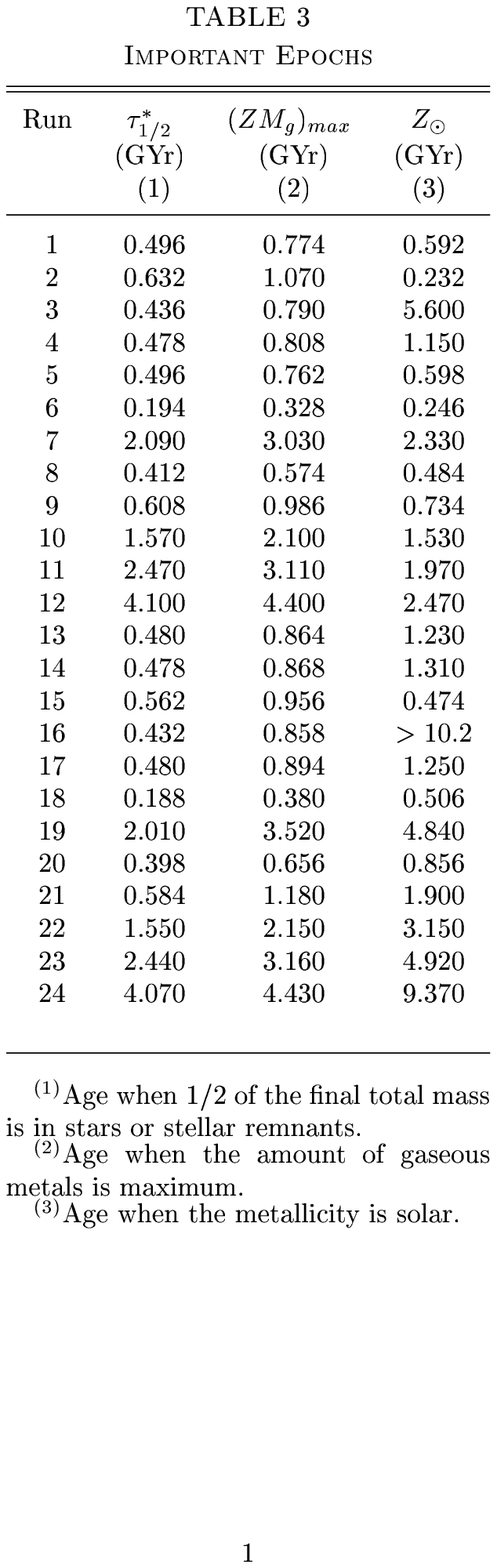}

%% file: figs.tex
%figures for chem.tex (Chemical evolution chapter)
\newpage
\setcounter{figure}{0}
%fig 1
\begin{figure}
  \includegraphics{f01.ps} \vspace*{4in}
\caption{ }
\end{figure}
\newpage
%fig 2
\begin{figure}
  \includegraphics{f02.ps} \vspace*{4in}
\caption{ }
\end{figure}
\newpage
%fig 3
\begin{figure}
  \includegraphics{f03.ps} \vspace*{4in}
\caption{ }
\end{figure}
%fig 4
\newpage
\begin{figure}
  \includegraphics{f04.ps} \vspace*{5.5in}
\caption{ }
\end{figure}
%fig5
\newpage
\begin{figure}
  \includegraphics{f05.ps} \vspace*{5.5in}
\caption{ }
\end{figure}
%fig6
\newpage
\begin{figure}
  \includegraphics{f06.ps} \vspace*{5.5in}
\caption{ }
\end{figure}
%fig7
\newpage
\begin{figure}
  \includegraphics{f07.ps} \vspace*{5.5in}
\caption{ }
\end{figure}
%fig8
\newpage
\begin{figure}
  \includegraphics{f08.ps} \vspace*{5.5in}
\caption{ }
\end{figure}
%fig 9
\begin{figure}
  \includegraphics{f09.ps} \vspace*{4in}
\caption{ }
\end{figure}
%fig 10
\clearpage
\begin{figure}
  \includegraphics{f10.ps} \vspace*{4.in}
\caption{ }

\end{figure}
%fig11
\begin{figure}
  \includegraphics{f11.ps} \vspace*{6.5in}
\caption{ }
\end{figure}
%fig12
\begin{figure}
  \includegraphics{f12.ps} \vspace*{6.5in}
\caption{ }
\end{figure}
%fig13
\begin{figure}
  \includegraphics{f13.ps} \vspace*{5.5in}
\caption{ }
\end{figure}
%fig14
\begin{figure}
  \includegraphics{f14.ps} \vspace*{5.5in}
\caption{ }
\end{figure}
%fig15
\begin{figure}
  \includegraphics{f15.ps} \vspace*{5.5in}
\caption{ }
\end{figure}
\newpage
%fig16
\begin{figure}
  \includegraphics{f16.ps} \vspace*{5.5in}
\caption{ }
\end{figure}
\newpage
%fig17
\begin{figure}
  \includegraphics{f17.ps} \vspace*{5.5in}
\caption{ }
\end{figure}